\documentclass[11pt]{article}
\usepackage{amsmath,amsthm,bm, amscd}
\usepackage{fullpage}
\usepackage[nofullpage,full,titlepage,nousetoc,nouselot,nouselof,hylinks,final]{boaz}
\usepackage{algorithm}
\usepackage{algorithmicx}
\usepackage{color}
\usepackage[colorlinks=true,citecolor=blue,linkcolor=blue]{hyperref}
\usepackage{graphicx}

\newcommand{\supp}{\mathsf{supp}}

\newcommand{\iext}{\mathsf{IExt}}

\newcommand{\zo}{\bits}

\newcommand{\acb}{\mathsf{AdvCB}}
\newcommand{\affcb}{\mathsf{AffineAdvCB}}

\newcommand{\snmExt}{\textnormal{snmExt}}

\newcommand{\A}{\mathcal{A}}
\newcommand{\Q}{\mathcal{Q}}
\newcommand{\scirc}{\hspace{0.1cm}\circ \hspace{0.1cm}}
\newcommand{\X}{\mathbf{X}}

\newcommand{\Y}{\mathbf{Y}}
\newcommand{\U}{\mathbf{U}}
\newcommand{\snm}{\mathsf{NM}}

\newcommand{\same}{\textnormal{$same^{\star}$}}
\newcommand{\cpy}{\textnormal{copy}}

\def\calX{{\mathcal X}}
\def\calY{{\mathcal Y}}

\DeclareMathOperator{\expect}{E}

\newcommand{\samp}{\mathsf{Samp}}
\newcommand{\anm}{\mathsf{anmExt}}

\theoremstyle{definition}

\newcommand{\eps}{\epsilon}

\newcommand{\Cond}{\mathsf{Cond}}

\newcommand{\aext}{\mathsf{AExt}}

\newcommand{\czext}{\mathsf{CZExt}}

\newcommand{\Supp}{\mathsf{Supp}}
\newcommand{\Raz}{\mathsf{Raz}}

\newcommand{\Ext}{\mathsf{Ext}}

\newcommand{\nmExt}{\mathsf{nmExt}}
\newcommand{\mac}{\mathsf{MAC}}
\newcommand{\bip}{\mathsf{IP}}
\newcommand{\Enc}{\mathsf{Enc}}

\newcommand{\TExt}{\mathsf{TExt}}
\newcommand{\itext}{\mathsf{ITExt}}
\newcommand{\sExt}{\mathsf{SumsetExt}}
\newcommand{\sext}{\mathsf{SpExt}}
\newcommand{\zuc}{\Cond}

\newcommand{\advsrcond}{\mathsf{advSRcond}}
\newcommand{\nmlext}{\mathsf{nmLExt}}

\newcommand{\BI}{\begin{itemize}}
\newcommand{\EI}{\end{itemize}}
\newcommand{\BE}{\begin{enumerate}}
\newcommand{\EE}{\end{enumerate}}

\newtheorem{thm}{Theorem}      % A counter for Theorems etc
\newcommand{\BT}{\begin{theorem}}   \newcommand{\ET}{\end{theorem}}
\newcommand{\BD}{\begin{definition}}   \newcommand{\ED}{\end{definition}}
\newcommand{\BCR}{\begin{corollary}} \newcommand{\ECR}{\end{corollary}}
\newtheorem{constr}[thm]{Construction}
\newcommand{\BCT}{\begin{constr}} \newcommand{\ECT}{\end{constr}}
\newcommand{\BL}{\begin{lemma}}   \newcommand{\EL}{\end{lemma}}

\newcommand{\BP}{\begin{proposition}}   \newcommand{\EP}{\end{proposition}}
\newcommand{\BCM}{\begin{claim}}   \newcommand{\ECM}{\end{claim}}
\newcommand{\BF}{\begin{fact}}   \newcommand{\EF}{\end{fact}}
\newcommand{\BA}{\begin{assumption}}   \newcommand{\EA}{\end{assumption}}

 \def\Del{{\Delta}}

\def\eps{\varepsilon}

 \def\ge{\geqslant}

\makeatletter
\def\ExtendSymbol#1#2#3#4#5{\ext@arrow 0099{\arrowfill@#1#2#3}{#4}{#5}}
\def\RightExtendSymbol#1#2#3#4#5{\ext@arrow 0359{\arrowfill@#1#2#3}{#4}{#5}}
\def\LeftExtendSymbol#1#2#3#4#5{\ext@arrow 6095{\arrowfill@#1#2#3}{#4}{#5}}
\makeatother

\newcommand{\hinf}{H_\infty}
\newcommand{\thinf}{\widetilde{H}_\infty}

\title{Two Source Extractors for Asymptotically Optimal Entropy, and (Many) More}

\author{Xin Li\thanks{
Department of Computer Science, Johns Hopkins University, \texttt{lixints@cs.jhu.edu}. Supported by NSF CAREER Award CCF-1845349 and NSF Award CCF-2127575.
}
}

\date{}
\begin{document}

\maketitle
\begin{abstract}
A long line of work in the past two decades or so established close connections between several different pseudorandom objects and applications, including seeded or seedless non-malleable extractors, two source extractors, (bipartite) Ramsey graphs, privacy amplification protocols with an active adversary, non-malleable codes and many more. These connections essentially show that an asymptotically optimal construction of one central object will lead to asymptotically optimal solutions to all the others. However, despite considerable effort, previous works can get close but still lack one final step to achieve truly asymptotically optimal constructions.

In this paper we provide the last missing link, thus simultaneously achieving explicit, asymptotically optimal constructions and solutions for various well studied extractors and applications, that have been the subjects of long lines of research.\ Our results include:

\begin{itemize}
    \item Asymptotically optimal seeded non-malleable extractors, which in turn give two source extractors for asymptotically optimal min-entropy of $O(\log n)$, explicit constructions of $K$-Ramsey graphs on $N$ vertices with $K=\log^{O(1)} N$, and truly optimal privacy amplification protocols with an active adversary.
    \item Two source non-malleable extractors and affine non-malleable extractors for some linear min-entropy with exponentially small error, which in turn give the first explicit construction of non-malleable codes against $2$-split state tampering and affine tampering with constant rate and \emph{exponentially} small error.
  
    \item Explicit extractors for affine sources, sumset sources, interleaved sources, and small space sources that achieve asymptotically optimal min-entropy of $O(\log n)$ or $2s+O(\log n)$ (for space $s$ sources).
    \item An explicit function that requires strongly linear read once branching programs of size $2^{n-O(\log n)}$, which is optimal up to the constant in $O(\cdot)$. Previously, even for standard read once branching programs, the best known size lower bound for an explicit function is $2^{n-O(\log^2 n)}$.
\end{itemize}

\end{abstract}

\thispagestyle{empty}
\newpage
\tableofcontents
\thispagestyle{empty}
%\end{titlepage}
\newpage
\pagenumbering{arabic}

\section{Introduction}
This paper studies a wide range of pseudorandom objects and applications. We first briefly survey each of them, and then state our main results.

\paragraph{Randomness Extractors.} Through decades of study, \emph{randomness extractors} have become fundamental objects in the area of pseudorandomness, with intimate connections to other areas such as cryptography, complexity theory, combinatorics and graph theory, and so on. The original motivation of randomness extractors comes from bridging the gap between uniform random strings required in many applications, and poor quality random sources available in practice. We use the following standard definition, where the \emph{min-entropy} of a random variable~$X$ is defined as $H_\infty(X)=\min_{x \in \supp(X)}\log_2(1/\Pr[X=x])$. For $X \in \zo^n$, we call $X$ an $(n,H_\infty(X))$-source, or an $H_\infty(X)$-source when $n$ is clear from context, and we say $X$ has \emph{entropy rate} $H_\infty(X)/n$. %of entropy and weak random source.
\iffalse
\begin{definition}
The \emph{min-entropy} of a random variable~$X$ is
\[ H_\infty(X)=\min_{x \in \supp(X)}\log_2(1/\Pr[X=x]).\]
For $X \in \zo^n$, we call $X$ an $(n,H_\infty(X))$-source, and we say $X$ has
\emph{entropy rate} $H_\infty(X)/n$.
\end{definition}
\fi

The goal is to extract almost uniform random bits from weak random sources. Unfortunately, no deterministic extractor can exist when the input is a single general weak random source even with min-entropy $k=n-1$. Hence, the study of randomness extractors has been focusing on several relaxed models. For example, Nisan and Zuckerman \cite{NisanZ96} introduced the notion of \emph{seeded extractors}, where the extractor has access to an additional independent short uniform random seed. Typically, we require the seeded extractor to be \emph{strong} in the sense that the output of the extractor is close to uniform even conditioned on the seed. It can be shown that there exist strong seeded extractors with excellent parameters, and we now have almost optimal constructions (e.g., \cite{LuRVW03, GuruswamiUV09, DvirW08, DvirKSS09}) after a long line of research. 

Although seeded extractors have proven to be quite useful, in certain applications (e.g., cryptography) even the short uniform random seed is undesirable, thus another relaxed model is to put more restrictions on the weak source, and construct \emph{deterministic} or \emph{seedless} extractors for a certain class of weak sources. We have the following definition. 
\BD
Let $\calX$ be a family of distribution over $\zo^n$. A function $\Ext : \zo^n \to \zo^m$ is a deterministic extractor for $\calX$ with error $\e$ if for every distribution $X \in \calX$, we have
\[ \Ext(X) \approx_{\e} U_m,\]
where $U_m$ stands for the uniform distribution over $\zo^m$, and $\approx_{\e}$ means $\e$ close in statistical distance. We say $\Ext$ is explicit if it is computable by a polynomial-time algorithm.
\ED
Historically, the most well studied class of sources is the class of two (or more) independent sources.\ Here, a simple probabilistic argument shows that there exist two source extractors for $(n, k)$ sources with $k=\log n+O(1)$, which is optimal up to the constant $O(1)$; and the first explicit construction of two source extractors was given by Chor and Goldreich \cite{ChorG88} more than 35 years ago, which achieves $k> n/2$. Due to their connections to explicit Ramsey graphs, and applications in distributed computing and cryptography with general weak random sources \cite{KalaiLRZ08, KalaiLR09}, such extractors have also been the subject of extensive study \cite{ChorG88, BarakIW04, BarakKSSW05, Raz05, Bourgain05, Rao06, BarakRSW06, Li11b, Li12b, Li13a, Li13b, Li15, Cohen15, DBLP:conf/stoc/Cohen16, CZ15, Li16, Coh16b, CL16, Coh16, BDT16, Cohen16, DBLP:conf/stoc/Cohen17, Li17, Li19, lewko_2019}. The ultimate goal is to construct explicit two source extractors for $k=\log n+O(1)$, which would also imply an (strongly) explicit Ramsey graph on $N$ vertices with no clique or independent set of size $O(\log N)$, solving a long standing open problem proposed by Erd\H{o}s \cite{erdos:ramsey} in his seminal paper that inaugurated the probabilistic method. Previously, the best explicit construction of two source extractors in terms of entropy is that of \cite{Li19}, which achieves $k=O(\log n \cdot \frac{ \log \log n}{\log \log \log n})$ and gives an explicit Ramsey graph on $N$ vertices with no clique or independent set of size $(\log N)^{O(\frac{\log \log \log N}{\log \log \log \log N})}$.

Deterministic extractors for many other classes of sources have been studied. These include for example bit fixing sources \cite{CGHFRS85, KZ06, GRS06, Rao09}, which are sources that are obtained by fixing some unknown bits of a uniform random string; affine sources \cite{GR08,Bourgain07,Rao09,Yehudayoff11,BK12,Shatiel11b,Li11a,Li16, ChattopadhyayGL21}, which generalize bit-fixing sources and are the uniform distributions over some unknown affine subspaces of a vector space;  samplable sources \cite{TV00, Viola14}, which are sources that are generated by small circuits or efficient algorithms; interleaved sources \cite{RY08, CZ15a}, which are a generalization of independent sources where the bits of the sources are mixed in some arbitrary order; and small-space sources \cite{KRVZ06}, where the sources are generated by a small width branching program. Deterministic extractors for these sources have applications in areas such as exposure-resilient cryptography \cite{CGHFRS85, KZ06}, Boolean circuit lower bounds \cite{DK11, FGHK15}, and best-partition communication complexity lower bound \cite{RY08}.

In \cite{ChattopadhyayL16}, Chattopadhyay and Li introduced the model of sumset sources, which is the sum of two (or more) independent weak random sources. This model generalizes many of the previously studied models, such as independent sources, bit fixing sources, affine sources, interleaved sources, and small space sources. For clarity we defer the formal definitions of these sources to later chapters. Thus, improved constructions of explicit extractors for sumset sources may also lead to improved explicit extractors for many of the above sources. While \cite{ChattopadhyayL16} only constructed explicit extractors for the sum of a constant number of $(n, k)$ sources with $k=\log^{O(1)} n$, a recent improvement by Chattopadhyay and Liao \cite{ChattopadhyayL22} gives explicit extractors for the sum of two independent $(n, k)$ sources with $k=O(\log n \log \log n \log \log \log^3 n)$. This in turn implies explicit extractors for affine sources and interleaved two sources with the same entropy. By an improved reduction from small space sources to sumset sources in \cite{ChattopadhyayL22}, this also gives explicit extractors for space $s$-sources with min-entropy $k=2s+O(\log n \log \log n \log \log \log^3 n)$. These are the previously best known constructions for each corresponding class of sources in terms of entropy.\footnote{We focus on affine sources over the field $\F_2$. For larger fields there are constructions with better parameters.} We note that non-explicitly, one can show that with high probability random functions are extractors for affine sources and interleaved two sources with entropy $k=O(\log n)$, and for space $s$-sources with min-entropy $k=2s+O(\log n)$. Interestingly, it is not clear if a random function is an extractor for the sum of two independent $(n, k)$ sources. However, since sumset sources are a generalization of two independent sources, the entropy lower bound of $\log n+O(1)$ for two source extractors also implies an entropy lower bound of $\log n/2+O(1)$ for the sum of two independent sources. 

\paragraph{Non-malleable extractors.} Motivated from cryptographic applications, an important variant of seeded/seedless extractors known as \emph{non-malleable extractors} has been the focus of much study in the past 15 years or so. Here, one or more inputs to the extractor are tampered with by an adversary, and the goal is to guarantee that the output of the extractor on the original inputs is still close to uniform even conditioned on the output of the extractor on the tampered inputs. To discuss non-malleable extractors, we start by defining tampering functions.

\begin{definition}[Tampering Function]
For any function $f:S \rightarrow S$, We say $f$ has no fixed points if $f(s) \neq s$ for all $s \in S$. For any $n>0$, let $\mathcal{F}_n$ denote the set of all functions $f: \{ 0,1\}^n \rightarrow \{0,1\}^n$. Any subset of $\mathcal{F}_n$ is a family of  tampering functions. 
\end{definition}

It is clear that if the tampering function is the identity function, then non-malleability is impossible.\ Thus, without loss of generality, for non-malleable extractors we only consider tampering functions with no fixed points (the more general definition is given in Definition~\ref{def:gnmext}). Depending on what the tampering function acts on, there are different models of non-malleable extractors. If the tampering acts on the seed of a seeded extractor, we get the notion of \emph{seeded non-malleable extractors}, introduced by Dodis and Wichs \cite{DW09}:

\begin{definition} [\cite{DW09}] A function $\snmExt:\{0,1\}^n \times \{ 0,1\}^d \rightarrow \{ 0,1\}^m$ is a strong seeded non-malleable extractor for min-entropy $k$ and error $\epsilon$ if the following holds:\ For any $(n, k)$ source $X$ and tampering function $\A : \{0,1\}^d \rightarrow \{0,1\}^d $ with no fixed points, we have
$$  \left |\snmExt(X,U_d) \scirc \snmExt(X,\A(U_d))  \scirc U_d- U_m \scirc  \snmExt(X,\A(U_d)) \scirc U_d \right | <\epsilon, $$where $U_m$ is independent of $U_d$ and $X$.
\end{definition}

Alternatively, if the tampering function acts on the inputs to a seedless extractor, then we get the notion of \emph{seedless non-malleable extractors}. This was first introduced by Cheraghchi and Guruswami \cite{CG14b} for the model of two independent sources:

\begin{definition}[\cite{CG14b}] %\label{def:t2}
A function $\nmExt : (\{ 0,1\}^{n})^C \rightarrow \{ 0,1\}^m$ is a $(k, \e)$-seedless non-malleable extractor for $C$ independent sources, if it satisfies the following property: Let $X_1, \cdots, X_C$ be $C$ independent $(n, k)$ sources, and $f_1, \cdots, f_C : \zo^n \to \zo^n$ be $C$ arbitrary tampering functions such that there exists an $f_i$ with no fixed points, then  $$ |\nmExt(X_1, \cdots, X_C) \circ \nmExt(f_1(X_1), \cdots, f_C(X_2)) - U_m \circ \nmExt(f_1(X_1), \cdots, f_C(X_2))| < \epsilon.$$ 
\end{definition}

Chattopadhyay and Li \cite{ChattopadhyayL17} adapted the definition to affine sources and affine tampering, thus leading to affine non-malleable extractors:

\BD [\cite{ChattopadhyayL17}]%\label{def:anmext}
A function $\anm:\zo^n \rightarrow \zo^m$ is a $(k, \e)$ affine non-malleable extractor if for any affine source $X$ with entropy at least $k$ and any affine function $f : \zo^n \rightarrow \zo^n$ with no fixed point, we have $$| \anm(X) \circ \anm(f(X)) - \U_m \circ \anm(f(X))| \leq \e.$$
\ED
%In addition, all the definitions can be naturally generalized to $t$-non-malleable extractors secure against $t$ different tampering functions.

Using the probabilistic method, one can prove the existence of all these non-malleable extractors with excellent parameters. For example, \cite{DW09} showed that seeded non-malleable extractors exist when $k>2m+2\log(1/\eps) + \log d + 6$ and $d>\log(n-k+1) + 2\log (1/\eps) + 5$. \cite{CG14b} showed that two source non-malleable extractors exist for $(n, k)$ sources when $k \geq m+\frac{3}{2}\log (1/\eps)+O(1)$ and $k \geq \log n+O(1)$. Similarly, it can be also shown that affine non-malleable extractors exist for entropy $k \geq 2m+2\log(1/\eps)+\log n+O(1)$.

However, constructing explicit non-malleable extractors turns out to be significantly harder than constructing standard extractors, despite considerable effort \cite{DLWZ11,CRS11,Li12a,Li12b, CGL15, Coh15nm, Coh16a, CL16, ChattopadhyayL17, Coh16, Cohen16, DBLP:conf/stoc/Cohen17, Li17, Li19}. Previously, the best explicit seeded non-malleable extractors are due to Li \cite{Li17, Li19}, which achieve $k \geq C(\log \log n+a \log(1/\e))$, $d=O(\log n)+\log(1/\e)2^{O(a(\log \log (1/\e))^\frac{1}{a})}$ and output length $\Omega(k)$, for some constant $C>1$ and any integer $a \in \N$; or $k \geq C(\log \log n+\log \log(1/\e) \log(1/\e))$ and $d =O(\log n+\log \log(1/\e) \log(1/\e))$ for some constant $C>1$. For two source non-malleable extractors, the best explicit constructions are due to Li \cite{Li19} and Chung, Obremski, Aggarwal \cite{ChungOA21}. The former achieves $k \geq (1-\gamma)n$ with error $2^{-\Omega(n \log \log n/\log n)}$ and output length $\Omega(n)$, for some constant $\gamma \in (0, 1)$; while the latter achieves $k_1 \geq (\frac{4}{5}+\gamma)n$ for the first source, $k_2 \geq C \log n$ for the second source, with some constants $C>1, \gamma \in (0, 1)$, error $2^{-\min(k_1, k_2)^{\Omega(1)}}$, and output length $\Omega(\min(k_1, k_2))$. The only known explicit affine non-malleable extractor is given in \cite{ChattopadhyayL17}, which achieves entropy $k \geq n-n^{\delta}$ for some constant $\delta \in (0, 1)$, error $2^{-n^{\Omega(1)}}$ and output length $n^{\Omega(1)}$.

\paragraph{Privacy amplification with an active adversary.} The basic problem of \emph{privacy amplification} was introduced by Bennett, Brassard, and Robert \cite{BennettBR88}. The situation arises where two parties with local (non-shared) uniform random bits aim to convert a shared secret weak random source $X$ into shared secret uniform random bits. This is achieved by a communication protocol, which is watched by an adversary with unlimited computational power.\ Such protocols are important in various applications such as quantum key distribution.\ While standard strong seeded extractors provide optimal one-round protocols for a passive adversary (i.e., an adversary who can only see the communications but cannot change them), they fail badly for an active adversary (i.e., an adversary who can arbitrarily change, delete and reorder messages). The main goal for the latter case is to design a protocol that uses as few number of interactions and as few bits of communications as possible, and achieves a shared uniform random string $R$ which is as long as possible.\ In this context, the difference between $H_{\infty}(X)$ and the length of the output is defined as the \emph{entropy loss}, together with a security parameter $s$, which ensures that the probability that any active adversary can successfully cause the two parties to output two different strings without being detected is at most $2^{-s}$. On the other hand, the two parties should achieve a shared secret string that is $2^{-s}$-close to uniform, if the adversary remains passive.\ We refer the reader to \cite{DLWZ11} for a formal definition.

A long line of work has been devoted to this problem \cite{MW97,dkrs,DW09,RW03,KR09,ckor,DLWZ11,CRS11,Li12a,Li12b,Li15b, CGL15, Coh15nm, Coh16a, CL16, Coh16, Cohen16, Li17, Li19}. In contrast to a passive adversary, here one round protocol can only exist when the entropy rate of $X$ is bigger than $1/2$, and the protocol has to incur a large entropy loss. For a source $X$ with entropy rate smaller than $1/2$, \cite{DW09} showed that any protocol needs at least two rounds with entropy loss at least $\Omega(s)$, and communication complexity at least $\Omega(\log n+s)$. Achieving a two-round protocol that asymptotically match these parameters for all possible security parameters $s$ is thus the ultimate goal (note that $s$ can be at most $\Omega(k)$ where $k=H_{\infty}(\X)$). Previously, the best known protocol is due to Li \cite{Li19}, which achieves two rounds with entropy loss $O(\log \log n+s)$, with communication complexity $O(\log n)+s2^{O(a(\log s)^\frac{1}{a})}$ for any constant integer $a \geq 2$ and $s$ up to $\Omega(k)$; or communication complexity $O(\log n+s \log^2 s)$ for $s$ up to $\Omega(k/ \log \log k)$. 

\paragraph{Non-malleable codes.} Non-malleable codes, introduced by Dziembowski, Pietrzak and Wichs \cite{DPW10}, are a generalization of standard error correcting codes to handle much larger classes of tampering. Informally, such a code is defined with respect to a specific family of tampering functions $\cal F$. The code consists of a randomized encoding function $E$ and a deterministic decoding function $D$, such that on any modified codeword $f(E(x))$ obtained from some function $f \in \cal F$ and some message $x$, the decoded message $x'=D(f(E(x)))$ is either the original message $x$, or $\e$-close to a completely unrelated message. The formal definition is given in Section~\ref{sec:nmcodes}. \cite{DPW10} shows that non-malleable codes have applications in tamper-resilient cryptography, and most notably, they can provide security guarantees even if the adversary can completely overwrite the codeword. 

Even with this relaxation, it can be seen that no non-malleable codes can exist if $\cal F$ is completely unrestricted. However, such codes do exist for many broad families of tampering functions. By now the study of non-malleable codes has grown into a large field with numerous publications, and we only survey some of the most related previous works here. One of the most natural and well studied families of tampering functions is the so called \emph{split-state} model, where a $k$-bit message $x$ is encoded into $t$ parts of messages $y_1, \cdots, y_t$, each of length $n$, so the rate of the code is $k/(tn)$. The adversary is then allowed to arbitrarily tamper with each $y_i$ independently. %In this case, the rate of the code is defined as $k/(tn)$. 

This model arises in many natural applications, for example when the $y_i$'s are stored in different parts of memory. Non-malleable codes in this model are also used in various non-malleable secret sharing schemes \cite{GK18}.\ Obviously, the case of $t=1$ corresponds to unrestricted tampering functions, and it is not possible to construct non-malleable codes. Thus the case of $t=2$ is the most general and interesting setting.  \cite{DPW10} first proved the existence of non-malleable codes in the split-state model, while Cheraghchi and Guruswami \cite{CG14a} showed that the optimal rate of non-malleable codes in the $2$-split-state model is $1/2$. Following a long line of research \cite{DKO13, ADL14, Agw14, ADKO15, CZ14, CGL15, Li17, KOS17, GMW18, Li19, AggarwalO20, AggarwalKOOS22}, Li \cite{Li19} gave the first explicit construction in the $2$-split-state model with constant rate and constant error $\e$, while Aggarwal and Obremski \cite{AggarwalO20} improved the error to be negligible $\e=2^{-k^{\Omega(1)}}$. The current best construction is due to \cite{AggarwalKOOS22}, which achieves rate $1/3$ and error $\e=2^{-k/\log^3 k}$.  

In \cite{ChattopadhyayL17}, Chattopadhyay and Li studied the model where the tampering function is any arbitrary affine function on the entire codeword (instead of acting on $2$ parts of the codeword independently). They give an explicit non-malleable code with rate $k^{-\Omega(1)}$ and error $2^{-k^{\Omega(1)}}$, which remains the best known construction to date.

\paragraph{Hardness against read-once linear branching program.} Branching programs are natural models to measure the space complexity of computation. A standard branching program is  a directed acyclic graph with one source and two sinks (labeled by $1$ and $0$), where each non-sink node is marked with an index of an input bit and has out-degree $2$. One outgoing edge is labeled by $0$ and the other is labeled by $1$. For any input, the computation of the branching program follows the natural path from the source to one sink, by reading the corresponding bits and going through the corresponding edges, and the input is accepted if the path ends in the sink with label $1$. The size of the branching program is defined as the number of its nodes, which roughly corresponds to $2^{O(s)}$ for space $s$ computation.

Unfortunately, proving non-trivial size lower bounds of explicit functions for general branching programs (e.g., those that can separate $\mathsf{P}$ from $\mathsf{LOGSPACE}$) seems beyond the reach of current techniques, hence essentially almost all research has been focusing on restricted models. Among these, the most well studied model is that of \emph{read once branching program}, or ROBP for short. In this model, in any computational path, each bit of the input is read at most once. Non-explicitly, an optimal lower bound of size $\Theta(2^{n-\log n})$ is known \cite{AndreevBCR99}. Explicitly, several previous works gave exponential lower bounds \cite{Wegener88, Zak84, DBLP:conf/fct/Dunne85, DBLP:journals/tcs/Jukna88, DBLP:journals/tcs/KrauseMW91, Simon1992ANL, doi:10.1137/S0097539795290349, DBLP:journals/ipl/Gal97, DBLP:journals/ipl/BolligW98, AndreevBCR99, DBLP:journals/tcs/Kabanets03}. However, the best known lower bound for an explicit function, due to Andreev, Baskakov, Clementi and Rolim \cite{AndreevBCR99}, is only $2^{n-O(\log^2 n)}$, and the bound of $2^{n-O(\log n)}$ is only known for a function in $\mathsf{DTIME}(2^{O(\log^2 n)}) \cap \mathsf{P/poly}$.  

Recently, motivated by strengthening tree-like resolution refutation lower bounds and average case lower bounds for parity decision trees, Gryaznov, Pudl\'{a}k, and Talebanfard \cite{GryaznovPT22} introduced the model of read once linear branching programs (ROLBP for short), where the queries on each computational path are generalized to be linear functions. To enforce the read once property, \cite{GryaznovPT22} defined two kinds of ROLBPs: a \emph{strongly} ROLBP requires that at any node, the span of the linear queries on all paths leading to this node has no non-trivial intersection with the span of the linear queries on all paths starting from this node, while a \emph{weakly} ROLBP only requires that the linear query at any node is not in the span of the linear queries on all paths leading to this node. It can be seen that both kinds of ROLBPs are generalizations of standard ROBPs.

\cite{GryaznovPT22} gave an explicit function which requires strong ROLBPs of size $\Omega(2^{n/3})$, which was subsequently improved by Chattopadhyay and Liao \cite{ChattopadhyayL22b} to $2^{n-\log^{O(1)} n}$.\footnote{In fact, these results also give average-case hardness for strongly ROLBPs.}

\subsection{Our Results}
We improve all of the above results, achieving asymptotically optimal constructions in almost all cases (except seedless non-malleable extractors, and the error and output length of seedless extractors). We list our main results according to the order of the areas that appear in the introduction. 

\paragraph{Seedless extractors.} Our results for seedless extractors can be summarized as follows.
\BT \label{thm:2source}
For every constant $\e>0$ there exists a constant $c>1$ and an explicit extractor $\TExt:  \zo^{2n} \to \zo$ with error $\e$, for the interleaving of two independent $(n, k)$ sources such that $k \geq c \log n$.
\ET

\BT %\label{thm:sumsetext}
For every constant $\e>0$ there exists a constant $c>1$ and an explicit extractor $\sExt: \zo^n  \to \zo$  with error $\e$, for the sum of two independent $(n, k)$ sources such that $k \geq c \log n$, or an affine source on $n$ bits with entropy $k \geq c \log n$.
\ET

\BT
For every constant $\e>0$ there exists a constant $c>1$ such that for every $s>0$ there exits an explicit extractor $\sext: \zo^n  \to \zo$  with error $\e$, for space-$s$ sources on $n$ bits with min-entropy $k \geq 2s+c \log n$.
\ET

All of the above theorems achieve asymptotically optimal entropy in the corresponding models. In addition, Theorem~\ref{thm:2source} immediately gives the following corollary about explicit Ramsey graphs.

\BCR
There is a constant $c>1$ such that for every integer $N$ there exists a (strongly) explicit Ramsey graph on $N$ vertices with no clique or independent set of size $K=\log^c N$. 
\ECR

\paragraph{Non-malleable extractors.} Our results for non-malleable extractors are summarized as follows.

\BT %\label{thm:stnmext}
  For any constant $\gamma>0$ there is a constant $C>0$ such that for any $0< \e < 1$ with $k \geq C \log(d/\e)$ and $d =C \log(n/\e)$, there is an explicit strong seeded non-malleable extractor for $(n, k)$ sources with seed length $d$, error $\e$ and output length $\frac{(1-\gamma)k}{2}$.  
\ET

This theorem achieves asymptotically optimal parameters in all aspects. In fact, we can also extend it to the stronger notion of $t$-non-malleable seeded extractors. See Section~\ref{sec:privacy} for details. Next we have seedless non-malleable extractors.

\BT
   There exists a constant $C>1$ such that for any constant $0< \gamma< 1$ and $k \geq C \log n$, there exists an explicit construction of a $((\frac{2}{3}+\gamma)n, k, 2^{-\Omega(k)})$ two-source non-malleable extractor with output length $\Omega(k)$.
\ET

This theorem improves both constructions in \cite{Li19} and \cite{ChungOA21}.\ Specifically, like in \cite{ChungOA21}, we can also handle the case where the second source only has logarithmic min-entropy, while we improve the entropy rate of the first source from $4/5+\gamma$ in \cite{ChungOA21} and $1-\gamma$ in \cite{Li19} to $2/3+\gamma$. Simultaneously, the error is also improved to an optimal $2^{-\Omega(k)}$, from $2^{-k^{\Omega(1)}}$ in \cite{ChungOA21} and $2^{-\Omega(k \log \log k/\log k)}$ in \cite{Li19}. We note that for applications in non-malleable codes, we don't really need such small entropy (any linear entropy suffices), but such two source non-malleable extractors have applications in privacy amplification with tamperable memory, see \cite{ChungOA21} for details.

\BT %\label{thm:nmaext}
There exists a constant $0< \gamma< 1$ such that for any $n \in \N$, there exists an explicit construction of a $((1-\gamma)n, 2^{-\Omega(n)})$ affine non-malleable extractor with output length $\Omega(n)$.
\ET

\paragraph{Privacy amplification.} Combining our optimal seeded non-malleable extractor with the protocol in \cite{DW09}, we get the following theorem.

\BT
There exists a constant $0<\alpha<1$ such that for any $n, k \in \N$, there is an explicit two-round privacy amplification protocol in the presence of an active adversary, that achieves any security parameter $s \leq \alpha k$, entropy loss $O(\log \log n+s)$, and communication complexity $O(\log n+s)$.
\ET

Our two-round protocol achieves asymptotically optimal parameters in all aspects, for security parameter up to $s=\Omega(k)$. The $O(\log \log n)$ term  is the best possible if using the two-round protocol in \cite{DW09}. This follows from the use of a message authentication code (MAC) that authenticates the seed of a strong seeded extractor with security parameter $s$, which has at least $\Omega(\log n)$ bits. Thus the MAC requires a key of length at least $\log \log n+s$. See \cite{DW09} for more details.

\paragraph{Non-malleable codes.} Using our seedless non-malleable extractors, we also get new constructions of non-malleable codes.

\BT \label{thm:2splitcode}
For any $n \in \N$ there exists a non-malleable code with efficient encoding and decoding against $2$-split-state tampering, which has message length $k$, block length $2n$, rate $k/(2n)=\Omega(1)$ and error $2^{-\Omega(k)}$.
\ET

\BT \label{thm:affinecode}
For any $n \in \N$ there exists a non-malleable code with efficient encoding and decoding against affine tampering, which has message length $k$, block length $n$, rate $k/n=\Omega(1)$ and error $2^{-\Omega(k)}$.
\ET

Both theorems are asymptotically optimal.\ Theorem~\ref{thm:2splitcode} achieves a smaller constant rate than the rate $1/3$ construction in \cite{AggarwalKOOS22}, but improves the error from $2^{-k/\log^3 k}$ to $2^{-\Omega(k)}$.\ Theorem~\ref{thm:affinecode} significantly improves the construction in \cite{ChattopadhyayL17}, with rate only $k^{-\Omega(1)}$ and error $2^{-k^{\Omega(1)}}$.

\paragraph{Hardness against read once linear branching program.} Our sumset extractor directly gives a hard function for strongly ROLBPs (in fact with any constant average-case hardness). We have

\BT
There is an explicit function $\sExt: \zo^n \to \zo$ that requires strongly read once linear branching program of size $2^{n-O(\log n)}$.
\ET

Our result improves the results of $\Omega(2^{n/3})$ in \cite{GryaznovPT22} and $2^{n-\log^{O(1)} n}$ in \cite{ChattopadhyayL22b}. Clearly, it also gives the first explicit function that requires standard ROBPs of size $2^{n-O(\log n)}$, improving the previously best known result of $2^{n-O(\log^2 n)}$ in \cite{AndreevBCR99}. By the $\Theta(2^{n-\log n})$ bound for standard ROBPs \cite{AndreevBCR99}, our result is optimal up to the constant in $O(.)$.  We remark that our affine extractor also directly gives an asymptotically optimal $2^{n-O(\log n)}$ size lower bound for DNF circuits with a bottom layer of parity gates, by the result in \cite{CohenS16}.

\subsection{Overview of the Techniques}
Before explaining our new ideas, we first recall the connections and reductions established in previous works.\ This allows us to reduce all the problems to a couple of central pseudorandom objects.
\paragraph{Connections between different pseudorandom objects and applications.} Non-malleable extractors have direct motivations and applications in cryptography. For example, \cite{DW09} shows that an optimal seeded non-malleable extractor gives an optimal two-round privacy amplification protocol with an active adversary. Similarly, \cite{CG14a} and \cite{ChattopadhyayL17} show that good two-source and affine non-malleable extractors give non-malleable codes against $2$-split state tampering and affine tampering.\ The idea is simple: the encoding function is to uniformly sample a pre-image of the message under the extractor function, and the decoding function is the extractor itself. Reducing the average case error of the extractor to the worst case guarantee of the code blows up the error $\e$ to $2^m \e$ where $m$ is the output length of the extractor. Thus, to achieve a constant rate it is crucial to have an exponentially small error $\e=2^{-\Omega(n)}$, while it is enough to work for any linear entropy $k=\Omega(n)$. For hardness against strongly ROLBPs, \cite{ChattopadhyayL22b} observed that, just like a standard ROBP, if one conditions on an internal node, then the programs before and after this node correspond to two independent sources. Hence this reduces the question of finding a hard function to the question of constructing a good extractor for the sum of two independent sources.  

Yet, previous works also established more surprising, and unexpected connections between non-malleable extractors and standard seedless extractors, which have been the underlying source of most of the recent progress on extractor theory.\ Specifically, the first such connection was established between seeded non-malleable extractors and two-source (and more generally independent source) extractors by Li \cite{Li12a, Li13a, Li13b}, where he showed sufficiently good seeded non-malleable extractors imply improved two source extractors. Using techniques from non-malleable extractors, this has led to Li's construction of the first explicit extractor for three independent $(n, k)$ sources with $k \geq \log^{O(1)} n$, output length $\Omega(k)$ and error $2^{-k^{\Omega(1)}}$ \cite{Li13b}. The construction uses two sources to produce a \emph{somewhere random} source with $n^{O(1)}$ rows, such that there exist a large fraction of (almost) uniform rows, and these rows are almost $t$-wise independent for some $t=\log^{O(1)} n$. The third source is then used to extract random bits from this somewhere random source.

Chattopadhyay and Zuckerman \cite{CZ15} further formalized this connection, and brought in another key improvement by applying a \emph{resilient function} directly to the somewhere random source, thus giving the first two source extractor for $k \geq \log^{O(1)} n$ with error $n^{-\Omega(1)}$.\ Afterwards, a series of works \cite{Li16, Coh16b, CL16, Coh16, BDT16} improved the reduction and eventually, \cite{BDT16} establishes that an optimal seeded non-malleable extractor\footnote{More accurately, a seeded non-malleable extractor against multiple tampering.} would give a two source extractor for entropy $O(\log n)$. Later, Li \cite{Li17} further established a connection between two source non-malleable extractors and seeded non-malleable extractors, which roughly says the following: a two source non-malleable extractor for any constant (less than $1$) entropy rate with error $2^{-\Omega(n)}$ would give an optimal seeded non-malleable extractor. Again, it is crucial here to have an exponentially small error of $2^{-\Omega(n)}$, while the entropy rate can be any constant less than $1$.\ Finally, these connections have been roughly extended to extractors for the sum of two independent sources in \cite{ChattopadhyayL22b}.\footnote{\cite{ChattopadhyayL22b} actually reduces extractors for sumset sources to good \emph{correlation breakers}, which are building blocks in two-source non-malleable extractors. We ignore these technical details here.}\ In summary, by the established connections, all the problems can be reduced to constructing explicit two-source and affine non-malleable extractors for any constant (less than $1$) entropy rate with error $2^{-\Omega(n)}$. 

\paragraph{Our new ideas.} Most of the above connections have been known for a while, yet the goal of constructing two-source non-malleable extractors with error $2^{-\Omega(n)}$ has been elusive so far. Indeed, more and more sophisticated techniques were developed in \cite{CL16, Coh16, Cohen16, Coh16b, DBLP:conf/stoc/Cohen17, Li17, Li19}, only resulting in the construction in \cite{Li19} which achieves error $2^{-\Omega(n \log \log n/\log n)}$. The bottleneck comes from the fact that all these constructions are based on some kind of \emph{alternating extraction} using an advice string. To get error $\e$ the length of the advice string is provably at least $\log(1/\e)$, while the alternating extraction appears to need at least some growing function $f(\log(1/\e))$ number of steps, where each step needs at least $\log(1/\e)$ entropy. This result in a total entropy of $f(\log(1/\e))\log(1/\e)$. Since the total entropy is $<n$ and $f$ is a growing function, this falls short of achieving error $2^{-\Omega(n)}$.

Luckily, there is one previous work by Chattopadhyay and Zuckerman \cite{CZ14} which does achieve error $2^{-\Omega(n)}$. Their constsruction relies on techinques from additive combinatorics, and does not use alternating extraction. However, their construction ($\czext$ for short) only gives a non-malleable extractor that requires $10$ independent $(n, k)$ sources with $k \geq (1-\gamma)n$ for some constant $\gamma >0$. In addition, the tampering function has to act independently on each of the $10$ sources, thus it is not a prior clear that this can give us anything for two source non-malleable extractors. Nevertheless, this construction is our starting point to provide the last missing link in the complete picture.

Essentially, we show how to get some kind of independence from just one weak source and an arbitrary function tampering with this source. To illustrate the basic idea, it helps to start with the example where $X$ is a uniform random string over $\zo^n$, while $f: \zo^n \to \zo^n$ is any linear tampering function. Let us divide $X$ evenly into $\ell$ blocks $X=X_1 \circ \cdots \circ X_{\ell}$, where each $X_i$ has $m=n/\ell$ bits. Consider the tampered input $X'=f(X)=X'_1 \circ \cdots \circ X'_{\ell}$. It is easy to see that there are linear functions $\{f^{ij}\}_{i, j \in [\ell]}$ such that for any $i \in [\ell]$, $X'_i =\sum_{j \in [\ell]} f^{ij}(X_j)$. If for some $i \in [\ell]$ there exists a $j \in [\ell], j \neq i$ such that $H(f^{ij}(X_j)) \geq \delta m$ for any constant $\delta>0$, then since $X_i$ and $X_j$ are independent, we have $H(X_i \circ X'_i) \geq H(X_i)+H(f^{ij}(X_j)) \geq (1+\delta)m$. This implies that the conditional entropy $H(X_i | X'_i)$ is at least $(1+\delta)m-m=\delta m$. In this case, we can apply an affine extractor for any linear entropy in \cite{Bourgain07, Yehudayoff11, Li11a}, so that the output on $X_i$ is close to uniform conditioned on the output on $X'_i$. This already achieves some kind of non-malleable extractor.

On the other hand, if for any $i \in [\ell]$ and any $j \in [\ell], j \neq i$, we have $H(f^{ij}(X_j)) < \delta m$, then we can \emph{fix} all $f^{ij}(X_j)$ where $i \neq j$. Note that conditioned on this fixing, the $X_i$'s are still independent, and furthermore the fixing does not cause any $X_i$ to lose much entropy. Specifically, each $X_i$ still has entropy at least $(1-\ell \delta)m$. Most importantly, with this fixing, each $X'_i$ is now a deterministic function of $X_i$! Thus, as long as $\ell \delta$ is small, we have obtained $\ell$ independent weak sources $\{X_i\}$ with $\ell$ tampering functions acting on each $X_i$ independently. Taking $\ell=10$ for example, at this point we can apply the function $\czext$ to the $X_i$'s, and the output will again be close to uniform even conditioned on the output on the $X'_i$'s. Thus, if we combine the outputs in both cases, we get a somewhere random source with $\ell+1$ rows such that one row is close to uniform conditioned on the corresponding row in the tampered output. We call this a non-malleable somewhere random source. With this object, it is now relatively easy to finish our construction using existing techniques. 

In summary, the high level key new idea of our constructions can be roughly stated as the following result of dichotomy, which leads to a ``win-win" situation: divide a weak source $X$ with sufficiently high entropy into $\ell$ blocks $X=X_1 \circ \cdots \circ X_{\ell}$, and consider the tampered version $X'=f(X)=X'_1 \circ \cdots \circ X'_{\ell}$. Then either (1) (in the case where $f$ ``mixes" the $X_i$'s well) there exists an $i \in [\ell]$ such that $X_i|X'_i$ has large entropy, or (2) (in the case where $f$ doesn't mix the $X_i$'s well) $X_1 \circ \cdots \circ X_{\ell}$ can be viewed as independent sources and $f$ can be viewed as $\ell$ functions $f=g_1 \circ \cdots \circ g_{\ell}$ where each $g_i$ acts on $X_i$ independently. 

However, making this idea formally work requires non-trivial techniques in both the constructions and the analysis. We now explain more technical details below.

\paragraph{Affine non-malleable extractors.} The previous analysis about a uniform random string $X$ can be relatively easily adapted to a high entropy affine source with slight modifications. Specifically, given an affine source on $n$ bits with entropy $k=(1-\gamma) n$ for some small constant $\gamma>0$, we now divide it into say $\ell+1$ blocks $X=X_1 \circ \cdots \circ X_{\ell} \circ X_{\ell+1}$, where each $X_i$ for $i \in [\ell]$ has $3 \gamma n$ bits and $X_{\ell+1}$ has $(1-3 \gamma \ell) n$ bits. Since $\ell=10$ is a constant, we can choose a small constant $\gamma$ and make sure the size of $X_{\ell+1}$ is much larger than the $X_i$'s. The plan is to use $X_1 \circ \cdots \circ X_{\ell}$ to generate the non-malleable somewhere random source, and then use $X_{\ell+1}$ to extract random bits. However, one issue here is that $X_1 \circ \cdots \circ X_{\ell}$ may be the same as $X'_1 \circ \cdots \circ X'_{\ell}$, in which case it is impossible to generate the non-malleable somewhere random source. To fix this, as in previous works, we need to first generate a small advice string $\alpha$ from $X$ such that $\alpha \neq \alpha'$ with probability $1-2^{\Omega(n)}$, where $\alpha'$ is the advice string generated from $X'$. We also need to keep the entropy of $X$ and the structure of an affine source conditioned on the generation of the advice strings. This turns out to be even trickier than the case of two-source non-malleable extractors, and we end up using two more blocks from $X$ and an improved advice generator for affine tampering based on that in \cite{ChattopadhyayL17}. To explain our main ideas we ignore these technical issues here, and refer the reader to Section~\ref{sec:affinenm} for details.

Now assume that we have already generated the advice string $\alpha$, and $X$ still has entropy $(1-\gamma)n$. The blocks of $X$ are no longer independent in general, but we show it is a convex combination of independent sources. Specifically, we view $X$ as the uniform random string subject to $\gamma n$ affine constraints. Conditioned on the fixing of the corresponding part of each constraint in each block, all blocks become independent. We can now do the same analysis as before. If for some $i \in [\ell]$ there exists a $j \in [\ell+1], j \neq i$ such that $H(f^{ij}(X_j))$ is large, then $H(X_i | X'_i)$ is also large. Otherwise, we can fix all the $f^{ij}(X_j)$'s with $i \in [\ell], j \in [\ell+1]$ and $i \neq j$. Conditioned on this fixing, the $X_i$'s are still independent with high entropy, and now all the $X'_i$'s with $i \in [\ell]$ are deterministic functions of the $X_i$'s. Thus we can apply an affine extractor to each $X_i$ with $i \in [\ell]$ and apply $\czext$ to $\{X_i \circ \alpha\}_{i \in [\ell]}$ (the concatenation with $\alpha$ ensures no fixed points with high probability). Combining all the outputs, we get a non-malleable somewhere random source $R$ with a constant number of rows, where each row has $\Omega(n)$ bits with error $2^{-\Omega(n)}$.

Note that $R$ and the tampered version $R'$ are deterministic functions of $\{X_i\}_{i \in [\ell]}$ and $\{X'_i\}_{i \in [\ell]}$. As long as $X_{\ell+1}$ has large enough entropy compared to the total size of $\{X_i\}_{i \in [\ell]}$ and $\{X'_i\}_{i \in [\ell]}$, a standard argument shows that there is an affine source $A$ contained in $X_{\ell+1}$ which is independent of $\{X_i\}_{i \in [\ell]}$ and $\{X'_i\}_{i \in [\ell]}$, and one can use linear seeded extractors to do alternating extraction between $R$ and $X_{\ell+1}$ to break the correlations. Indeed we apply an \emph{affine correlation breaker}, such as those developed in \cite{Li16, ChattopadhyayL22} to $X_{\ell+1}$ and each row of $R$, using the index of the corresponding row as the advice string, and finally take the XOR of all outputs. We argue that the output is non-malleable as follows. Without loss of generality assume that the first row of $R$ (denoted by $R_1$) is close to uniform conditioned on the first row of $R'$ (denoted by $R'_1$). We first fix $R'_1$ and all the outputs produced in the affine correlation breaker with $X'_{\ell+1}$ and $R'_1$. By using linear seeded extractors appropriately and keeping the output length to be small, we can ensure that (1) the affine structure of the sources is preserved, (2) $A$ still has high entropy and is independent of $\{X_i\}_{i \in [\ell]}$ and $\{X'_i\}_{i \in [\ell]}$, and (3) $R_1$ is still close to uniform. Now the affine correlation breaker guarantees that the output from $(X_{\ell+1}, R_1)$ is close to uniform given all the other outputs from $(X_{\ell+1}, R)$ and $(X'_{\ell+1}, R')$. Therefore once we take the XOR of the outputs, the string produced from $X$ is close to uniform conditioned on the string produced from $X'$. The key point is that $R$ only has a constant number of rows, thus the index of each row only has a constant number of bits, and $R_1$ and $X_{\ell+1}$ has $\Omega(n)$ entropy. Hence, we can achieve error $2^{-\Omega(n)}$ with output length $\Omega(n)$.

\paragraph{Two-source non-malleable extractors.} The case of two-source non-malleable extractors is more complicated, as here we don't have the nice structure of affine sources. Again, we ignore the issue of generating advice strings, and assume that we are given an advice string $\alpha \in \zo^{\Omega(n)}$ such that $\alpha \neq \alpha'$ with probability $1-2^{\Omega(n)}$, where $\alpha'$ is the advice string generated from the tampered inputs. We refer the reader to Section~\ref{sec:tnmext} for details. 

We show how to use a single source and the advice string to generate a \emph{non-malleable somewhere high entropy source}, which is a source $R$ with a constant number of rows, each row with $\Omega(n)$ bits, and there exists a row $i$ such that $H_{\infty}(R_i|R'_i) \geq \Omega(n)$ (again $R'$ is the tampered version). We call this function a \emph{non-malleable somewhere condenser with advice}. This is similar in spirit to, and can be viewed as the non-malleable analogue of the reduction given in \cite{BarakKSSW05}, which shows how to turn an independent source extractor into a somewhere condenser, that converts any weak random source with any linear entropy into a constant number of rows such that one row has entropy rate $0.9$.

Specifically, given an $(n, k)$ source $X$ with $k \geq (1-\beta)n$ for some small constant $\beta>0$, let us again divide $X$ evenly into $\ell=10$ blocks $X=X_1 \circ \cdots \circ X_{\ell}$ where each $X_i$ has $m=n/\ell$ bits. The non-malleable somewhere condenser produces a random variable $R$ with $\ell+1$ rows, where for each $i \in [\ell]$, $R_i=X_i$, and $R_{\ell+1}=\czext(X_1 \circ \alpha,  \cdots, \circ X_{\ell} \circ \alpha)$.

The analysis is more subtle and relies on carefully dividing $X$ into a convex combination of subsources.\ Let $X'=X'_1 \circ \cdots \circ X'_{\ell}$ be the tampered input.\ Without loss of generality assume $X$ is the uniform distribution on a set $S \subseteq \zo^n$ with size $2^{(1-\beta)n}$.\ Similar to \cite{BarakKSSW05}, for each $i \in [\ell]$, we define $H_i$ to be the set which contains heavy elements in the support of $(X_i, X'_i)$, e.g., $H_i = \{(y, y') \in \zo^{2m}: \Pr[(X_i, X'_i) = (y, y')] \geq 2^{-(1+3\beta) m} \}$. We divide $S$ into two subsets: $S' = \{x \in S : \exists i, (x_i, x'_i) \notin H_i\}$ and $S'' = \{x \in S : \forall i, (x_i, x'_i) \in H_i\}= S \setminus S'$. If either $S'$ or $S''$ is small, e.g., has size at most $2^{(1-\beta)n-\beta m}$, then we can safely ignore it since it only has probability mass at most $2^{-\beta m}$. Otherwise we consider $S'$ and $S''$ separately, since $X$ is just a convex combination of the uniform distributions over $S'$ and $S''$.

$S'$ is relatively easy to handle. Given that $|S'| \geq 2^{(1-\beta)n-\beta m}$, if we divide $S'$ into disjoint subsets by grouping all $x \in S'$ with the same smallest index $i$ such that $(x_i, x'_i) \notin H_i$ together, then on average each subset has size roughly $2^{(1-\beta)n-\beta m}/\ell$. Since all elements in the subset are light elements, the uniform distribution over the subset has min-entropy at least $(1+3\beta) m-\beta m-\log \ell > (1+\beta)m$. This means that if we consider the subsource corresponding to the uniform distribution over each subset, then roughly $H_{\infty}(X_i|X'_i) \geq \beta m=\Omega(n)$.

Taking care of $S''$ is much trickier. In this case, we want to argue that somehow, $X_1, \cdots, X_{\ell}$ can be viewed as independent sources and the tampering function $f$ can be viewed as $f=g_1 \circ \cdots \circ g_{\ell}$ where each $g_i$ acts on $X_i$ independently. Note that in this case, for any $x \in S''$ and any $i \in [\ell]$, we have $(x_i, x'_i) \in H_i$. Our first step is to remove those elements $x \in S''$ such that there exists an $i \in [\ell]$ and too many $y' \in \zo^m$ (say $> 2^{\beta n+ 6\beta m}$ such $y'$'s) where $(x_i, y') \in H_i$. Intuitively, these are the strings where the tampering function $f$ mixes too much entropy from the blocks $\{X_j, j \neq i\}$ into $X'_i$, and thus are bad for our purpose. By definition of $H_i$, for any $i$ we have $|H_i| \leq 2^{(1+3\beta) m}$. Hence the number of such $x$'s cannot be too large, and is at most $\ell 2^{(1+3\beta) m}/2^{\beta n+ 6\beta m} \cdot 2^{(\ell-1)m} < 2^{(1-\beta)n-2\beta m}$. Thus, removing these strings only cause $X$ to lose probability mass at most $2^{-2\beta m}$.

Let $S^*$ be the subset of $S''$ after removing the bad strings. It is clear that $S^*$ still has a large size, i.e., $|S^*| \geq (1-2^{-\beta m})2^{(1-\beta)n- \beta m} > 2^{n-2\ell \beta m}$. We now consider $X^*$, the uniform distribution over $S^*$, and $X'^*=f(X^*)$. Let $S_i$ be the support of $X^*_i$. The large size of $S^*$ guarantees that each $S_i$ also has large size, in fact $|S_i| \geq 2^{(1-2 \ell \beta) m}$. We now consider the sources $(Y_1, Y_2, \cdots, Y_{\ell})$ where each $Y_i$ is the independent uniform distribution over $S_i$. To construct the functions $g_1, \cdots, g_{\ell}$, for any $y \in S_i$ we define the set $W^y_i=\{y' \in \bits^m: y \circ y' \in H_i\}$. Since we have removed the bad $x$'s, we now have $|W^y_i| \leq 2^{\beta n+ 6\beta m}$ for any $i$ and any $y \in S_i$. We now consider a \emph{random} function $g=(g^1, g^2, \cdots, g^{\ell})$ where for any $i \in [\ell]$ and any $y \in S_i$, let $g^i(y)$ be a random element independently uniformly chosen from $W^y_i$. For all other $y \in \bits^m$ let $g^i(y)=0^m$.

With the random functions, for any $x \in S^*$ we have $\Pr[(x, x')=(x, g(x))] \geq (2^{-\ell(\beta n+ 6\beta m)}) \geq 2^{-7 \ell \beta n}$ by the independence of the $g^i$'s. Now by linearity of expectation, there exists a subset $V \subseteq S^*$ with $|V| \geq 2^{-7 \ell \beta n} |S^*| \geq 2^{-O( \ell \beta n)} \Pi_{i \in [\ell]}|S_i|$ such that for any $x \in V$, $(x, x')=(x, g(x))$. We can now remove the set $V$ from $S^*$ and repeat the above process. As long as there are at least $2^{-\beta n}|S^*|$ strings left, the same argument will give us a new set $V \subseteq S^*$ with $|V| \geq 2^{-O( \ell \beta n)} \Pi_{i \in [\ell]}|S_i|$ and a new function $g=(g^1, g^2, \cdots, g^{\ell})$ such that for any $x \in V$, $(x, x')=(x, g(x))$. Repeat this process until there are less than $2^{-\beta n}|S^*|$ strings left, and we have divided $S^*$ into large disjoint subsets $\{V_{q} \subseteq \bits^n, q \in \Q\}$ with $\ell$-split state tampering functions $\{g_q: (\bits^m)^{\ell} \to (\bits^m)^{\ell}, q \in \Q\}$, and a small subset left with less than $2^{-\beta n}|S^*|$ strings.  

Observe that $X^*$ is $2^{-\beta n}$-close to a convex combination of the uniform distributions on $\{V_{q}, q \in \Q\}$, while each subset $V_q$ has  large density in the set $\Pi_{i \in [\ell]} S_i$. Since each $S_i$ itself is large, with an appropriate choice of parameters, we can ensure that for any $q\in \Q$, $\czext(Y_1 \circ \alpha, Y_2 \circ \alpha, \cdots, Y_{\ell} \circ \alpha)$ is close to uniform conditioned on $\czext(g_q(Y_1) \circ \alpha', g_q(Y_2) \circ \alpha', \cdots, g_q(Y_{\ell}) \circ \alpha')$. We then show by Lemma~\ref{lem:condition1} that conditioned on the event $(Y_1, Y_2, \cdots, Y_{\ell}) \in V_q$, $\czext(Y_1 \circ \alpha, Y_2 \circ \alpha, \cdots, Y_{\ell} \circ \alpha)$ is close to having min-entropy $\Omega(n)$ conditioned on $\czext(g_q(Y_1) \circ \alpha', g_q(Y_2) \circ \alpha', \cdots, g_q(Y_{\ell}) \circ \alpha')$. This takes care of $S''$.

Ignoring the error (which is $2^{-\Omega(n)}$) and the issue of convex combination of subsources, we have now obtained a non-malleable somewhere condenser.\ The rest of the construction and analysis is relatively straightforward. In the actual construction, we will divide $X$ into more blocks, for example $X=X_1 \circ \cdots \circ X_{\ell} \circ X_{\ell+1}$ where each $X_i$ has $\Omega(n)$ bits, but $X_{\ell+1}$ has much larger size compared to the previous blocks. We use $(X_1,  \cdots, X_{\ell})$ to obtain the non-malleable somewhere high entropy source with a constant number of rows. Then, using sum-product theorem based condensers in \cite{BarakKSSW05, Raz05, Zuc07}, we can boost the conditional min-entropy rate from $\Omega(1)$ to $0.9$, while only increasing the number of rows by a constant factor. At this point we apply an extractor by Raz \cite{Raz05} to each row and the second source $Y$, which effectively converts the non-malleable somewhere high entropy source into a non-malleable somewhere random source. Fix $(X_1,  \cdots, X_{\ell})$ and $(X'_1,  \cdots, X'_{\ell})$, we argue that $X$ and $Y$ are still independent, and $X_{\ell+1}$ has enough entropy left. We can now use the non-malleable somewhere random source and a standard correlation breaker to extract uniform random bits from $X_{\ell+1}$, thus achieving a two-source non-malleable extractor by a similar argument as that of the affine non-malleable extractor. Again, the key point is that the somewhere random source only has a constant number of rows, and each row and $X_{\ell+1}$ has $\Omega(n)$ entropy. Hence, we can achieve error $2^{-\Omega(n)}$ with output length $\Omega(n)$.

The above gives a two-source non-malleable extractor for entropy rate $1-\beta$ with some small constant $\beta>0$. We can decrease the entropy of the first source to $k_1 \geq (2/3+\gamma)n$ and the entropy of the second source to $k_2 \geq O(\log n)$ by first taking a slice of the first source with size $n/3$, then applying the sum-product theorem based condensers in \cite{BarakKSSW05, Raz05, Zuc07}, Raz's extractor \cite{Raz05} to the second source, and a strong seeded extractor (e.g., those in \cite{GuruswamiUV09}) to the first source to boost the entropy rate. This will result in a constant number of rows in both sources such that there exists one row where both sources have very high entropy rate. We can then apply the advice generator, our new two-source non-malleable extractor for entropy rate $1-\beta$, and finally the correlation breaker and taking the XOR of the outputs. See Section~\ref{sec:tnmext} for details. 

\paragraph{Efficiently sampling the pre-image.} For applications in non-malleable codes, we need to design efficient algorithms to sample uniformly from the pre-image of any output of our seedless non-malleable extractors.\ Thus we appropriately modify our extractors, roughly following the same approach as in \cite{Li17}. However, to achieve error $2^{-\Omega(n)}$, we can no longer use a Reed-Solomon code in the advice generator, since this only achieves error $2^{-\Omega(n/\log n)}$. Instead, we use an asymptotically good linear binary code whose dual code is also asymptotically good. This implies that for some constant $\eta>0$, any $\eta$ fraction of columns in the generator matrix are linearly independent. %with the following property: any submatrix of the generator matrix with large enough size has full rank. We show that such codes can be constructed from small biased sample space (e.g., \cite{nn, Ta-Shma17}). 

\subsection{Organization of the Paper} The rest of the paper is organized as follows.\ In section~\ref{sec:prelim} we give some preliminaries and previous works we use.\ In section~\ref{sec:affinenm} we give our affine non-malleable extractor.\ In section~\ref{sec:nmswcond}, ~\ref{sec:advcb} and ~\ref{sec:tnmext} we give our non-malleable somewhere condenser, non-malleable correlation breaker, and two-source non-malleable extractor.\ In section~\ref{sec:app} we give various applications where most of them achieve asymptotically optimal parameters. We conclude with some open problems in section~\ref{sec:conc}.

%{Organization of the paper.}
\section{Preliminaries} \label{sec:prelim}
We use capital letters for random variables and corresponding small letters for their instantiations. We use letters with prime for the tampered version.\ Let $|S|$ denote the cardinality of the set~$S$.\ For $\ell$ a positive integer,
$U_\ell$ denotes the uniform distribution on $\zo^\ell$.\ When used as a component in a vector, each $U_\ell$ is independent of the other components.
%Let $\dbZ_r$ denote the cyclic group $\dbZ/(r\dbZ)$, and for $S$ a set, $U_S$ denotes the uniform distribution on $S$
%and let $\F_q$ denote the finite field of size $q$.
All logarithms are to the base 2.

\subsection{Probability Distributions}
\begin{definition} [statistical distance]Let $W$ and $Z$ be two distributions on
a set $S$. Their \emph{statistical distance} (variation distance) is
\begin{align*}
\Delta(W,Z) \eqdef \max_{T \subseteq S}(|W(T) - Z(T)|) = \frac{1}{2}
\sum_{s \in S}|W(s)-Z(s)|.
\end{align*}
\end{definition}

We say $W$ is $\eps$-close to $Z$, denoted $W \approx_\eps Z$, if $\Delta(W,Z) \leq \eps$.
For a distribution $D$ on a set $S$ and a function $h:S \to T$, let $h(D)$ denote the distribution on $T$ induced by choosing $x$ according to $D$ and outputting $h(x)$.
%We often view a distribution as a function whose value at a sample point is the probability of that sample point.
%Thus $\lone{W-Z}$ denotes the $\ell_1$ norm of the difference of the distributions specified by the random variables $W$ and $Z$, which equals $2\Delta(W,Z)$.

\BL \label{lem:sdis}
For any function $\alpha$ and two random variables $A, B$, $\Delta(\alpha(A), \alpha(B)) \leq \Delta(A, B)$.
\EL

\subsection{Somewhere Random Sources and Extractors}

\begin{definition} [Somewhere Random sources] \label{def:SR} A source $X= (X_1, \cdots, X_t)$ is $(t \times r)$
  \emph{somewhere-random} (SR-source for short) if each $X_i$ takes values in $\bits^r$ and there is an $i$ such that $X_i$ is uniformly distributed.
\end{definition}

\BD[subsource] 
Let $X$ be an $n$-bit source in some probability space.\ We say that an event $A$ is determined by $X$ if there exists a function $f : \zo^n \to \zo$ such that $A = \{f(X) =1\}$.\ We say $X_0$ is a subsource of $X$ if there exists an event $A$ that is determined by $X$ such that $X_0 = (X|A)$.
\ED
%\iffalse
\BD
An elementary somewhere-k-source is a vector of sources $(X_1, \cdots, X_t)$, where some $X_i$ is a $k$-source.\ A somewhere $k$-source is a convex combination of elementary somewhere-$k$-sources.
\ED

\BD
A function $C: \bits^n \times \bits^d \to \bits^m$ is a $(k \to l, \e)$-somewhere-condenser if for every $k$-source $X$, the vector $(C(X, y)_{y \in \bits^d})$ is $\e$-close to a somewhere-$l$-source. When convenient, we call $C$ a rate-$(k/n \to l/m, \e)$-somewhere-condenser.   
\ED
\iffalse
\BD
A function $C: \bits^n \times \bits^d \to \bits^m$ is a $(k \to l, \e)$-condenser if for every $k$-source $X$, $C(X, U_d)$ is $\e$-close to some $l$-source. When convenient, we call $C$ a rate-$(k/n \to l/m, \e)$-condenser.   
\ED

\BD (Block Sources)
A distribution $X=X_1 \circ X_2 \circ \cdots, \circ X_t$ is called a $(k_1, k_2, \cdots, k_t)$ block source if for all $i=1, \cdots, t$, we have that for all $x_1 \in \Supp(X_1), \cdots, x_{i-1} \in \Supp(X_{i-1})$, $H_{\infty}(X_i|X_1=x_1, \cdots, X_{i-1}=x_{i-1}) \geq k_i$, i.e., each block has high min-entropy even conditioned on any fixing of the previous blocks. If $k_1=k_2 = \cdots =k_t=k$, we say that $X$ is a $k$ block source.
\ED
\fi

\begin{definition}(Seeded Extractor)\label{def:strongext}
A function $\Ext : \bits^n \times \bits^d \rightarrow \bits^m$ is  a \emph{strong $(k,\eps)$-extractor} if for every source $X$ with min-entropy $k$
and independent $Y$ which is uniform on $\zo^d$,
\[ (\Ext(X, Y), Y) \approx_\eps (U_m, Y).\]
\end{definition}

%\iffalse
\begin{definition}

A function $\TExt : \bits^{n_1} \times \bits^{n_2} \rightarrow \bits^m$ is  a \emph{strong two source extractor} for min-entropy $k_1, k_2$ and error $\e$ if for every independent  $(n_1, k_1)$ source $X$ and $(n_2, k_2)$ source $Y$, 

\[ |(\TExt(X, Y), X)-(U_m, X)| < \e\]

and

\[ |(\TExt(X, Y), Y)-(U_m, Y)| < \e,\]
where $U_m$ is the uniform distribution on $m$ bits independent of $(X, Y)$. 
\end{definition}
%\fi

%\iffalse
\subsection{Average Conditional Min Entropy}
\label{avgcase}

%Dodis and Wichs originally defined non-malleable extractors with respect to average conditional min-entropy, a notion defined by
%Dodis, Ostrovsky, Reyzin, and Smith \cite{dors}.

\begin{definition}
The \emph{average conditional min-entropy} is defined as
\begin{align*}
 \thinf(X|W) &= - \log \left (\expect_{w \leftarrow W} \left [ \max_x \Pr[X=x|W=w] \right ] \right )
\\ &= - \log \left (\expect_{w \leftarrow W} \left [2^{-\hinf(X|W=w)} \right ] \right ).
\end{align*}
\end{definition}

\iffalse
Average conditional min-entropy tends to be useful for cryptographic applications.
By taking $W$ to be the empty string, we see that average conditional min-entropy is at least as strong as min-entropy.
In fact, the two are essentially equivalent, up to a small loss in parameters.

We have the following lemmas.
\fi

\begin{lemma} [\cite{dors}]
\label{entropies}
For any $s > 0$,
$\Pr_{w \leftarrow W} [\hinf(X|W=w) \geq \thinf(X|W) - s] \geq 1-2^{-s}$.
\end{lemma}

\BL [\cite{dors}] \label{lem:amentropy}
If a random variable $B$ has at most $2^{\ell}$ possible values, then $\thinf(A|B) \geq \hinf(A)-\ell$.
\EL

\subsection{Seedless Non-Malleable Extractors}

\BD[Seedless Non-Malleable Extractor] \label{def:gnmext}
\[
 \cpy(x,y) =
  \begin{cases}
   x & \text{if } x \neq \same \\
   y       & \text{if } x  = \same

  \end{cases}
\]

A function $\nmExt : \{ 0,1\}^{n} \rightarrow \{ 0,1\}^m$ is a $(k, \eps)$-seedless non-malleable extractor with respect to a class $\cal X$ of sources over $\zo^n$ and a class $\cal F$ of tampering functions acting on $\zo^n$, if for every $\X \in \cal X$ with min-entropy $k$ and every $f \in \cal F$, there is a distribution $\cal D$ over $\zo^m \cup \{\same\}$ such that for an independent $\Y$ sampled from $D$, we have
$$ (\nmExt(\X), \nmExt(f(\X))) \approx_{\eps}  (U_m, \cpy(\Y, U_m)),$$ 
where the second $U_m$ is the same random variable as the first one.
\ED

The following is a generalization of the connection shown by Cheraghchi and Guruswami \cite{CG14b}. 

\begin{thm}\label{thm:connection} Let $\nmExt: \{0,1\}^{n} \rightarrow \{0,1\}^{m}$  be a polynomial time computable seedless non-malleable extractor that works for min-entropy $n$ with error $\epsilon$ with respect to a class of tampering functions $\F$ acting on $\zo^n$. Further suppose there is a sampling algorithm $\samp$ that on any input $z \in \zo^m$ runs in time $\poly(n)$ and samples from a distribution that is $\epsilon'$-close to uniform on the set $\nmExt^{-1}(s)$.  

Then there exists an efficient construction of a non-malleable code  with respect to the tampering family $\F$ with block length $=n$, relative rate  $\frac{m}{n}$ and error $2^{m}\epsilon+\epsilon'$.
\end{thm}

The non-malleable code  is define in the following way: For any message $s \in \{ 0,1\}^m$, the encoder of the non-malleable code outputs $\samp(s)$. For any codeword $c \in \{0,1\}^{n}$, the decoder outputs $\nmExt(c)$. 

In this paper we will mainly consider the classes of $2$-split state tampering and affine tampering.

\subsection{Linear Error Correcting Codes}
\BD
An $[n, k, d]$ code $C$ is a dimension $k$ linear subspace of the vector space $\F^n_2$, such that any non-zero codeword in $C$ has Hamming weight (the number of $1$'s) at least $d$. Thus, the encoding function for any message $x \in \F^k_2$ is $y = x G$ for some matrix $G \in \F^{k \times n}_2$. We say $G$ is the generator matrix of $C$, and $C$ is explicit if $G$ can be constructed in time $\poly(n)$. The dual code of $C$, $C^{\perp}$, is defined to be the linear subspace of $\F^n_2$ orthogonal to $C$, i.e., $C^{\perp}=\{z \in \F^n_2 : \forall y \in C, \langle z, y \rangle = 0\}$.
\ED

\subsection{Prerequisites from Previous Work}
\iffalse
One-time message authentication codes (MACs) use a shared random key to authenticate a message in the information-theoretic setting.
\begin{definition} \label{def:mac}
A function family $\{\mac_R : \bits^{d} \to \bits^{v} \}$ is a $\e$-secure one-time MAC for messages of length $d$ with tags of length $v$ if for any $w \in \bits^{d}$ and any function (adversary) $A : \bits^{v} \to \bits^{d} \times \bits^{v}$,

\[\Pr_R[\mac_R(W')=T' \wedge W' \neq w \mid (W', T')=A(\mac_R(w))] \leq \e,\]
where $R$ is the uniform distribution over the key space $\bits^{\ell}$.
\end{definition}

\begin{theorem} [\cite{kr:agree-close}] \label{thm:mac}
For any message length $d$ and tag length $v$,
there exists an efficient family of $(\lceil  \frac{d}{v} \rceil 2^{-v})$-secure
$\mac$s with key length $\ell=2v$. In particular, this $\mac$ is $\eps$-secure when
$v = \log d + \log (1/\e)$.\\
More generally, this $\mac$ also enjoys the following security guarantee, even if Eve has partial information $E$ about its key $R$.
Let $(R, E)$ be any joint distribution.
Then, for all attackers $A_1$ and $A_2$,

\begin{align*}
\Pr_{(R, E)} [&\mac_R(W')=T' \wedge W' \neq W \mid W = A_1(E), \\ &~(W', T') = A_2(\mac_R(W), E)] \leq \left \lceil  \frac{d}{v} \right \rceil 2^{v-\thinf(R|E)}.
\end{align*}

(In the special case when $R\equiv U_{2v}$ and independent of $E$, we get the original bound.)
\end{theorem}

\begin{remark}
Note that the above theorem indicates that the MAC works even if the key $R$ has average conditional min-entropy rate $>1/2$.
\end{remark}
\fi

Sometimes it is convenient to talk about average case seeded extractors, where the source $X$ has average conditional min-entropy $\thinf(X|Z) \geq k$ and the output of the extractor should be uniform given $Z$ as well. The following lemma is proved in \cite{dors}.

\BL \cite{dors} \label{lem:avext}
For any $\delta>0$, if $\Ext$ is a $(k, \e)$ extractor then it is also a $(k+\log(1/\delta), \e+\delta)$ average case extractor.
\EL

For a strong seeded extractor with optimal parameters, we use the following extractor constructed in \cite{GuruswamiUV09}.

\BT [\cite{GuruswamiUV09}] \label{thm:optext} 
For every constant $\alpha>0$, and all positive integers $n,k$ and any $\e>0$, there is an explicit construction of a strong $(k,\e)$-extractor $\Ext: \bits^n \times \bits^d \to \bits^m$ with $d=O(\log n +\log (1/\e))$ and $m \geq (1-\alpha) k$. In addition, for any $\e> 2^{-k/3}$ this gives a strong $(k, \e)$ average case extractor with $m \geq k/2$.%It is also a strong $(k, \e)$ average case extractor with $m \geq (1-\alpha) k-O(\log n+\log (1/\e))$.
\ET

We need the following ``invertible" extrator from \cite{Li17}.
\BT [\cite{Li17}] \label{thm:iext}
There exists a constant $0<\alpha<1$ such that for any $n \in \N$ and $2^{-\alpha n}< \e<1 $ there exists a linear seeded strong extractor $\iext: \bits^n \times \bits^d \to \bits^{0.3 d}$ with $d=O(\log(n/\e))$ and the following property. If $X$ is a $(n,0.9n)$ source and $R$ is an independent uniform seed on $\{ 0,1\}^{d}$, then $$ |(\iext(X,R),R) - (U_{0.3 d},R)| \leq \e.$$ 
Furthermore for any $s \in \{ 0,1\}^{0.3 d}$ and any $r \in  \{ 0,1\}^{d}$, $| \iext(\cdot,r)^{-1}(s)|= 2^{n-0.3 d}$.
\ET

We will also use a sampler in our construction.

\BD[Averaging sampler \cite{Vadhan04}]\label{def:samp} A function $\samp: \{0,1\}^{r} \rightarrow [n]^{t}$ is a $(\mu,\theta,\gamma)$ averaging sampler if for every function $f:[n] \rightarrow [0,1]$ with average value $\frac{1}{n}\sum_{i}f(i) \ge \mu$, it holds that 
$$ \Pr_{i_1,\ldots,i_t \leftarrow \samp(U_{R})}\left [  \frac{1}{t}\sum_{i}f(i) < \mu - \theta \right ] \leq \gamma.$$
$\samp$ has distinct samples if for every $x \in \{ 0,1\}^{r}$, the samples produced by $\samp(x)$ are all distinct.
\ED

\BT [\cite{Vadhan04}] \label{thm:samp} Let $1 \geq \delta \geq 3\tau > 0$. Suppose that $\samp: \zo^r \to [n]^t$ is an $(\mu,\theta,\gamma)$ averaging sampler with distinct samples for $\mu=(\delta-2\tau)/\log(1/\tau)$ and $\theta=\tau/\log(1/\tau)$. Then for every $\delta n$-source $X$ on $\zo^n$, the random variable $(U_r,  X_{Samp(U_r)})$ is $(\gamma+2^{-\Omega(\tau n)})$-close to $(U_r, W)$ where for every $a \in \zo^r$, the random variable $W|_{U_r=a}$ is $(\delta-3\tau)t$-source.
\ET

\BT[\cite{Vadhan04}]\label{thm:sampler} For every $0< \theta< \mu<1$, $\gamma>0$, and $n \in \N$, there is an explicit $(\mu,\theta,\gamma)$ averaging sampler $\samp: \zo^r \to [n]^t$
 that uses
 \begin{itemize}
 \item $t$ distinct samples for any $t \in [t_0, n]$, where $t_0=O(\frac{1}{\theta^2} \log(1/\gamma))$, and
 \item $r=\log (n/t)+\log(1/\gamma)\poly(1/\theta)$ random bits.
 \end{itemize}
\ET

\BT [\cite{ChorG88}] \label{thm:ip}
For every $0<m< n$ there is an explicit two-source extractor $\bip: \bits^n \times \bits^n \to \bits^m$ based on the inner product function, such that if $X, Y$ are two independent $(n, k_1)$ and $(n, k_2)$ sources respectively, then

\[(\bip(X, Y), X) \approx_{\e} (U_m, X) \text{ and } (\bip(X, Y), Y) \approx_{\e} (U_m, Y),\]
where $\e=2^{-\frac{k_1+k_2-n-m-1}{2}}.$
\ET

\BT [\cite{Raz05}] \label{thm:Razext}
For any $n_1, n_2, k_1, k_2, m$ and any $0 < \delta < 1/2$ with
\begin{itemize}
    \item $n_1 \geq 6 \log n_1 + 2 \log n_2$
    \item $k_1 \geq (0.5 + \delta)n_1 + 3 \log n_1 + \log n_2$
    \item $k_2 \geq 5 \log(n_1 - k_1)$
    \item $m \leq \delta \min[n_1/8, k_2/40]-1$
\end{itemize}

There is a polynomial time computable strong 2-source extractor $\Raz: \bits^{n_1} \times \bits^{n_2} \to \bits^{m}$ for min-entropy $k_1, k_2$ with error $2^{-1.5m}$.
\ET

\BT [\cite{BarakKSSW05, Raz05, Zuc07}] \label{thm:swcondenser}
For any constant $\beta, \delta>0$, there is an efficient family of rate-$(\delta \to 1-\beta, \e=2^{-\Omega(n)})$-somewhere condensers $\zuc: \bits^n \to (\bits^m)^D$ where $D=O(1)$ and $m=\Omega(n)$. 

\ET

We need the following explicit construction of seedless non-malleable extractors in \cite{CZ14}.

\BT\label{thm:czext}
There exists a constant $\gamma>0$ and an explicit $(k, \e)$-seedless non-malleable extractor for $10$ independent sources $\czext: (\bits^n)^{10} \to \bits^m$ with $k=(1-\gamma)n$, $\e=2^{-\Omega(n)}$ and $m=\Omega(n)$.
\ET

The following standard lemma about conditional min-entropy is implicit in \cite{NisanZ96} and explicit in \cite{MW97}.

\begin{lemma}[\cite{MW97}] \label{lem:condition} 
Let $X$ and $Y$ be random variables and let ${\calY}$ denote the range of $Y$. Then for all $\e>0$, one has
\[\Pr_Y \left [ H_{\infty}(X|Y=y) \geq H_{\infty}(X)-\log|{\calY}|-\log \left( \frac{1}{\e} \right )\right ] \geq 1-\e.\]
\end{lemma}

\begin{lemma}[\cite{Zuc07}] \label{lem:ksource} 
The statistical distance of a random variable $X$ to the closest $k$-source is $\sum_s \max(X(s)-2^{-k}, 0)$.
\end{lemma}

We need the following lemma.

\BL \label{lem:condition1}
Let $X, Y, X'$ be random variables such that $X, X'$ have the same support, $X'$ has min-entropy $k$ and is independent of $Y$, and $(X, Y) \approx_{2^{-r}} (X', Y)$. Let ${\calY}$ denote the range of $Y$. Let $E$ be an event such that $\Pr[E] \geq 2^{-t}$.  Then for all $\e>0$, $(X, Y)|E$ is $\e+2^{t-r}$-close to another distribution $(\Tilde{X}, \Tilde{Y})$, such that for every $y \in \supp(\Tilde{Y})$, $\Tilde{X}|\Tilde{Y}=y$ is a $k-t-\log(1/\e)$ source. 
\EL

\begin{proof}
    For any $x \in \supp(X)$ and $y \in \supp(Y)$, let $\Del_{x,y}=|\Pr[X=x, Y=y]-\Pr[X'=x, Y=y]|$. Thus we have $\sum_{x \in \supp(X), y \in \supp(Y)} \Del_{x,y} \leq 2^{-r}$, and $\Pr[X=x, Y=y] \leq \Pr[X'=x, Y=y]+\Del_{x,y} \leq 2^{-k}\Pr[Y=y]+\Del_{x,y}$. 
    Then 
        \[\Pr[X=x|(Y=y, E)]=\frac{\Pr[X=x, Y=y, E]}{\Pr[Y=y, E]} \leq \frac{\Pr[X=x, Y=y]}{\Pr[Y=y, E]}\]

    Define the following set: $B=\{y \in {\calY}: \Pr[Y=y |E] < \e \Pr[Y=y]\}$. %and $B_2=\{y \in \bits^m: \Pr[Y=y |E] < \e \cdot 2^{-m}\}$.

    We have 
    $$\Pr[(Y|E) \in B]=\sum_{y \in B}\Pr[Y=y |E] < \sum_{y \in B}\e\Pr[Y=y] \leq \e.$$ 
    
   Whenever $(Y|E) \notin B$, we have $\Pr[Y=y |E] \geq \e \Pr[Y=y]$. Therefore for any $x \in \supp(X)$,

\begin{align*}
   \Pr[X=x|(Y=y, E)] & \leq \frac{\Pr[X=x, Y=y]}{\Pr[Y=y, E]} \leq \frac{2^{-k}\Pr[Y=y]+\Del_{x,y}}{\Pr[Y=y, E]} \\
    & = \frac{2^{-k}\Pr[Y=y]}{\Pr[Y=y|E]\Pr[E]}+\frac{\Del_{x,y}}{\Pr[Y=y|E]\Pr[E]} \\
    & \leq 2^{-k+t}/\e+\frac{2^t\Del_{x,y}}{\Pr[Y=y|E]}.
\end{align*}

Thus by Lemma~\ref{lem:ksource}, there exists a $k-t-\log(1/\e)$ source $Z_y$ such that the statistical distance of $X|(Y=y, E)$ to $Z_y$ is at most $\sum_{x \in \supp(X)} \frac{2^t\Del_{x,y}}{\Pr[Y=y|E]}$. Now let $(\Tilde{X}, \Tilde{Y})$ be the following distribution: first sample $\Tilde{Y}=y$ according to $Y|(E, Y \notin B)$, and then sample $\Tilde{X}$ as $Z_y$. Notice that this distribution is $\e+\sum_{x \in \supp(X), y \in {\calY} \setminus B}\frac{2^t\Del_{x,y}}{\Pr[Y=y|E]}\cdot {\Pr[Y=y|E]}\leq \e+2^{t-r}$-close to $(X, Y)|E$. On the other hand, for every $y \in \supp(\Tilde{Y})$, conditioned on $\Tilde{Y}=y$, $\Tilde{X}$ is a $k-t-\log(1/\e)$ source. 
\end{proof}

We also need the following lemma.

\BL \label{lem:jerror}\cite{Li13b}
Let $(X, Y)$ be a joint distribution such that $X$ has range $\calX$ and $Y$ has range $\calY$. Assume that there is another random variable $X'$ with the same range as $X$ such that $|X-X'| = \e$. Then there exists a joint distribution $(X', Y)$ such that $|(X, Y)-(X', Y)| = \e$.
\EL

%\BT [\cite{nn, Ta-Shma17}] \label{thm:smallbias}
%For any $\e>0$ there exists an explicit construction of $\e$-biased sample space for $n$ random variables with size $O(\frac{n}{\e^{2+o(1)}})$.
%\ET

We need the following theorem due to Guruswami \cite{10.1007/s00037-009-0281-5}, about binary linear codes such that both the code and its dual are asymptotically good.

\BT \label{thm:dcode} \cite{10.1007/s00037-009-0281-5}
For every integer $i \geq 1$ there is a $[n_i, n_i/2, d_i]$ code where $n_i = 42 \cdot 8^{i+1}$ and $d_i \geq n_i/30$. Moreover, the generator matrix can be constructed in $poly(n_i)$ time, and the dual of this linear code is a $[n_i, n_i/2, d'_i]$ code where $d'_i \geq n_i/30$.
\ET

This immediately gives the following theorem.

\BT \label{thm:bcode}
For any $n \in \N$ there is an explicit construction of the generator matrix of an $[n', n, d]$ code with $n'=O(n), d=\Omega(n)$ that satisfies the following property: any $d$ columns in the generator matrix are linearly independent.
\ET

%\begin{proof}
%  We take an explicit construction of a binary self-dual $[n', k'=n'/2, d]$ code. By the results in \cite{904540, DOUGHERTY201014}, such codes can be constructed efficiently for any even length $n'$. Thus, we can choose $k'=n$ and $n'=2n$. Furthermore, it is well known \cite{MACWILLIAMS1972153} that any binary self-dual code is asymptotically good, hence we have $d=\Omega(n')=\Omega(n)$.  
%\end{proof}
\section{Affine Non-Malleable Extractor} \label{sec:affinenm}
We use this section to construct affine non-malleable extractors. First we define affine sources and recall the definition of affine non-malleable extractors.

\BD (affine source over $\F_2$) A distribution $X$ over $\F_2^n$ is an $(n, k)$ affine source if $X$ is the uniform distribution over some affine subspace of $\F_2^n$ with dimension $k$.
\ED

\BD \label{def:anmext}
A function $\anm:\zo^n \rightarrow \zo^m$ is a $(k, \e)$ affine non-malleable extractor if for any affine source $X$ with entropy at least $k$ and any affine function $f : \zo^n \rightarrow \zo^n$ with no fixed point, we have $$| \anm(X),\anm(f(X)) - \U_m, \anm(f(X))| \leq \e.$$
\ED

We need the following definitions and lemmas about correlation breakers.

\BD [Correlation breaker with advice] \cite{CGL15, Coh15nm} \label{def:advcb}A function

\[\acb: \bits^n \times \bits^{n'} \times \bits^a \to \bits^m\] is called a $(k, k', \eps)$-correlation breaker with advice if the following holds. Let $X, X'$ be $n$-bit random variables with $H_{\infty}(X) \geq k$, $Y, Y'$ be $n'$-bit
random variables with $H_{\infty}(Y) \geq k'$, such that $(X, X')$ is independent of $(Y, Y')$. Then, for any pair of distinct $a$-bit strings $\alpha, \alpha'$,

\[(\acb(X,Y,\alpha), \acb(X',Y',\alpha')) \approx_{\eps} (U_m,\acb(X',Y',\alpha')).\] 
In addition, we say that $\acb$ is strong if
\begin{align*}
&(\acb(X,Y,\alpha), \acb(X',Y',\alpha'), Y, Y')  \approx_{\eps} (U_m,\acb(X',Y',\alpha'), Y, Y').
\end{align*}
\ED

The following definition generalizes the definition of affine correlation breakers in \cite{ChattopadhyayL22}.

\BD \label{def:affacb}
A function $\affcb : \bits^n \times \bits^d \times \bits^a \to \bits^m$ is a $t$-affine correlation breaker for entropy $k$ with error $\e$ (or a $(t, k, \e)$-affine correlation breaker for short) if for every distributions $X, X_1, \cdots, X_t, A, A_1, \cdots, A_t, B, B_1, \cdots, B_t \in \bits^n$, $Y, Y_1, \cdots, Y_t \in \bits^d$ and strings $\alpha, \alpha_1, \cdots, \alpha_t \in \bits^a$ such that

\begin{itemize}
    \item $X = A+B$, and for any $i \in [t]$, $X_i = A_i+B_i$,
    \item $H_{\infty}(A) \geq k$ and $Y$ is uniform,
    \item $(A, A_1, \cdots, A_t)$ is independent of $(B,  B_1, \cdots, B_t, Y, Y_1, \cdots, Y_t)$,
    \item $\forall i \in [t]$, $\alpha \neq \alpha_i$,
\end{itemize}
it holds that
$$(\affcb(X, Y, \alpha), \{\affcb(X_i, Y_i, \alpha_i)\}_{i \in [t]}) \approx_{\e} (U_m, \{\affcb(X_i, Y_i, \alpha_i)\}_{i \in [t]}).$$

We say $\affcb$ has degree $t$, and $\affcb$ is strong if
$$(\affcb(X, Y, \alpha), \{\affcb(X_i, Y_i, \alpha_i), Y_i\}_{i \in [t]}) \approx_{\e} (U_m, \{\affcb(X_i, Y_i, \alpha_i), Y_i\}_{i \in [t]}).$$
\ED

The following theorem can be proved by using essentially the same proof as in \cite{ChattopadhyayL22} for the special case of $X=X_1= \cdots =X_t$.

\BT [\cite{ChattopadhyayL22}] \label{thm:affcb} 
Let $C$ be a large enough constant. Suppose that there exists an explicit $(d_0, d_0, \e)$-strong correlation breaker with advice $\acb : \bits^n \times \bits^{d_0}
\times \bits^a \to \bits^{C \log^2 (t+1) \log(n/\e)}$ for some $n, t \in \N$. Then there exists an explicit strong $t$-affine correlation breaker $\affcb :\bits^n \times \bits^d \times \bits^a \to \bits^m$ with error $O(t \e)$ for entropy $k=O(td_0+tm+t^2\log(n/\e))$, where $d=O(td_0+m+t\log^3(t+1)\log(n/\e))$.
\ET

To apply this transformation, we use a standard correlation breaker with advice from \cite{Li19}.

\BT [\cite{Li19}] \label{thm:scb} 
There exists an explicit (standard) $(d, d, \e)$ correlation breaker with advice $\bits^n \times \bits^d \times \bits^a \to \bits^m$, where $d = O(m + \log(n/\e) \cdot \frac{\log (a)}{\log \log (a)})$.
\ET

Combining the above two theorems we have the following theorem.

\BT \label{thm:acbmain} 
For any $t \in \N$ there exists an explicit strong $t$-affine correlation breaker $\affcb :\bits^n \times \bits^d \times \bits^a \to \bits^m$ with error $O(t \e)$ for entropy $k=O(tm+t\log(n/\e) \cdot \frac{\log (a)}{\log \log (a)}+t^2\log(n/\e))$, where $d=O(tm+t\log(n/\e) \cdot \frac{\log (a)}{\log \log (a)}+t\log^3(t+1)\log(n/\e))$.
\ET

We also need the following affine extractor.

\BT [\cite{Bourgain07, Li11a, Yehudayoff11}]\label{thm:aext}
For any constant $\delta >0$ there is an explicit affine extractor $\aext: \bits^n \to \bits^m$ for entropy $k \geq \delta n$, with $m=\Omega(n)$ and error $2^{-\Omega(n)}$.
\ET

\begin{lemma}[Affine Conditioning~\cite{Li11a}]
    \label{lem:affinecond}
    Let $X$ be any affine source on $\{0,1\}^n$. Let $L:\{0,1\}^n \to \{0,1\}^m$ be any affine function. Then there exist independent affine sources $A,B$ such that:
    \begin{itemize}
        \item $X = A + B$
        \item For every $b\in \Supp(B),\; L(b) = c$ for some $c\in \{0,1\}^m$.
        \item $H(A) = H(L(A))$ and there exists an affine function $L^{-1}: \{0,1\}^m \to \{0,1\}^n$ such that $A = L^{-1}(L(A))$.
    \end{itemize}
\end{lemma}

\iffalse
\BT \label{thm:bcode}
For any constant $\delta>0$ there is an explicit construction of the generator matrix of an $[n', n, d]$ code with $n'=O(n), d=\Omega(n)$ that satisfies the following property: For any $t \in \N$ with $(1/2+\delta)n' \leq t \leq n$, any $t \times t$ submatrix has full rank.
\ET

\begin{proof}
  We take an explicit construction of an $\e$-biased sample space with $n$ binary random variables and $\e=\delta/2$. Now let the generator matrix be such that each row corresponds to a random variable, and each column corresponds to a point in the sample space. By Theorem~\ref{thm:smallbias}, we have $n'=O(\frac{n}{\e^{2+o(1)}})=O(n) > 3n$ and $d \geq (1/2-\e/2)n'=\Omega(n)$.

Now consider any $t \times t$ submatrix with $(1/2+\delta)n' \leq t \leq n$.  Since any non-trivial linear combination of these rows in the original matrix has at most $(1/2+\delta/2)n'$ $0$'s. The restriction to the $t$ columns cannot be $0$. Hence the $t \times t$ submatrix has rank $t$. 
\end{proof}

%\iffalse
We use the following theorem proved by Chattopadhyay and Zuckerman \cite{CZ14}.
\BT [\cite{CZ14}]\label{thm:czext}
There is a constant $0<\gamma<1$ and an explicit non-malleable $10$-source extractor $\czext$ for $(n, (1-\gamma)n)$ sources with error $2^{-\Omega(n)}$ and output length $\Omega(n)$.
\ET
\fi
\subsection{The Extractor Construction}
Our affine non-malleable extractor is given below.

\begin{algorithm}[H]
    \caption{$\anm(x)$}
    \label{alg:anmext}
    \begin{algorithmic}
        \medskip
        \State \textbf{Input:} $x \in \bits^{n}$ --- two $n$ bit strings.
        \State \textbf{Output:} $w \in \bits^m$ --- a string with length $m=\Omega(n)$.
        \\\hrulefill 
        \State \textbf{Sub-Routines and Parameters: } \\
        \item Let $0<\gamma< \alpha <1/1000$ be two constants to be chosen later.
        \item Let $\aext$ be the affine extractor from Theorem~\ref{thm:aext}.
       \item Let $\iext$ be the invertible linear seeded extractor form Theorem~\ref{thm:iext}.
       \item Let $\Enc$ be the encoding function of the linear code from Theorem~\ref{thm:bcode}.
              \item Let $\samp$ be the average sampler from Theorem~\ref{thm:sampler}.
       \item Let $\czext$ be the non-malleable extractor from Theorem~\ref{thm:czext}.
       \item Let $\affcb$ be the $t$-affine correlation breaker with advice from Theorem~\ref{thm:acbmain}, for $t=20$.      
        \\\hrulefill \\
        
        %Set $x^0=x$ and let $i=0$. Initially $x^i$ has only $n_0 = 1$ row. 
         \begin{enumerate}
        \item Divide $x$ into $x=x_0 \circ x_1 \circ x_2$, where $x_0, x_1$ each has $2 \gamma n$ bits, and $x_2$ has $n'=(1-4 \gamma)n$ bits. 

\item Compute $z_0=\aext(x_0)$ and $z_1=\aext(x_1)$, each outputs $\Omega(\gamma n)$ bits.

\item Let $\overline{x}=\Enc(x_2)$. For each $i=0, 1$, use $z_i$ and $\samp$ to sample $s=\Omega(n) \leq \gamma n$ distinct bits from $\overline{x}$, let the resulted string be $\widetilde{x_i}$. %Similarly, Use $z_1$ and $\samp$ to sample $s=\Omega(n) \leq \gamma n$ distinct bits from $\overline{x}$, let the resulted string be $\tilde{x_1}$.

\item Let $\widetilde{\alpha}=x_0 \circ x_1 \circ \widetilde{x_0} \circ \widetilde{x_1}$. Divide $x_2$ into $x_3 \circ x_4 \circ \cdots \circ x_{13} \circ x_{14} \circ \hat{x}$ such that $x_i$ has $\alpha n$ bits for any $3 \leq i \leq 12$, $x_{13}$ has $30\alpha n$ bits, $x_{14}$ has $100\alpha n$ bits, while $\hat{x}$ has $n-140 \alpha n-4 \gamma n \geq 2n/3$ bits. 

\item For each $i \in [10]$, compute $v_i=\aext(x_{i+2})$ with $\delta=\gamma$ in Theorem~\ref{thm:aext}. Compute $v_{11} = \czext(x_3 \circ \widetilde{\alpha}, \cdots, x_{12} \circ \widetilde{\alpha})$. All outputs will have $\Omega(n) \leq \gamma n$ bits. 

\item For each $i \in [11]$, compute $r_i=\affcb(x_{13}, v_i, i)$ with degree $t=20$, which outputs $s=\Omega(n) \leq \gamma n$ bits, and $r=\oplus_{i \in [11]} r_i$.

\item Finally compute $w=\iext(x_{14}, r)$ which outputs $s=\Omega(n) \leq \gamma n$ bits.
        \end{enumerate}

    \end{algorithmic}
\end{algorithm}

To analyze the algorithm we first have the following lemma.

\BL \label{lem:affindep}
Let $X$ be an affine source over $n$ bits with entropy $n-r$, and $X=X_1 \circ \cdots \circ X_t$ where each $X_i$ has $n_i$ bits, so $\sum_i n_i =n$. Then $X$ is a convex combination of affine sources $X^j$, where for each $j$ and $X^j=X^j_1 \circ \cdots \circ X^j_t$, the $\{X^j_i\}_{i \in [t]}$'s are independent affine sources, and each $X^j_i$ has entropy at least $n_i-r$.
\EL

\begin{proof}
    We view $X=X_1 \circ \cdots \circ X_t$ as the uniform random string over $\bits^n$, subject to $r$ affine constraints. Each constraint corresponds to a linear equation with the bits of $X$, thus for each $X_i$ we can fix the corresponding linear part within $X_i$ to a specific bit. Conditioned on these fixings, the $X_i$'s are still independent, and each of them is an affine source with entropy at least $n_i-r$.
\end{proof}

We now have the following theorem.

\BT \label{thm:nmaext}
There exists a constant $0< \gamma< 1$ such that for any $n \in \N$, there exists an explicit construction of a $((1-\gamma)n, 2^{-\Omega(n)})$ affine non-malleable extractor with output length $\Omega(n)$.
\ET

\begin{proof}
We use capital letters with prime to denote the corresponding random variables produced from the tampered input. By Lemma~\ref{lem:affindep}, without loss of generality we can assume $X=X_0 \circ X_1 \circ X_3 \circ \cdots \circ X_{13} \circ X_{14} \circ \hat{X}$, where each part is an independent affine source with entropy deficiency $\gamma n$. In particular, $X_0$ and $X_1$ both have entropy at least $\gamma n$.

We now argue that $\widetilde{\alpha} \neq \widetilde{\alpha}'$ with high probability. First note that if $X_0 \neq X'_0$ or $X_1 \neq X'_1$, then we are done. Otherwise, we must have $X_2 \neq X'_2$. Note that $X'_2$ is an affine function of $X=X_0 \circ X_1 \circ X_2$. Let $L_0, L_1: \bits^{2\gamma n} \to \bits^{(1-4\gamma)n}$ be the affine functions that correspond to the contributions of $X_0, X_1$ in $X'_2$, respectively. We now have two cases.

\begin{description}
\item [Case 1.] $H(L_0(X_0)) \leq \gamma n/2$. We fix $X_1, X_2$, and $L(X_0)$, and conditioned on this fixing, $X_0=X'_0$ still has entropy at least $\gamma n/2$. Therefore by Theorem~\ref{thm:aext}, $Z_0 \approx_{2^{-\Omega(n)}} U_{\Omega(n)}$. Note that 
\[\Enc(X_2)+\Enc(X_2')=\Enc(X_2+X_2'),\]

and under the fixings, $X_2+X_2'$ is also fixed to be a non-zero string. Therefore, by Theorem~\ref{thm:bcode} and Theorem~\ref{thm:sampler}, $\widetilde{X_0} \neq \widetilde{X_0}'$ with probability $1-2^{-\Omega(n)}$ over the further fixing of $X_0$.

\item [Case 2.] $H(L_0(X_0)) > \gamma n/2$. We fix $X_2$, and conditioned on this fixing, $X'_2=L_0(X_0)+L_1(X_1)+a$ for some $a \in \bits^{(1-4\gamma)n}$. Notice that 
\[\Enc(X_2)+\Enc(X_2')=\Enc(X_2+X_2')=\Enc(L_0(X_0))+\Enc(L_1(X_1))+b,\]

for some $b \in \bits^{O(n)}$. Therefore, we have 

\[\widetilde{X_1} + \widetilde{X_1}' = \samp(\Enc(L_0(X_0)), Z_1)+\samp(\Enc(L_1(X_1))+b, Z_1).\]

By Theorem~\ref{thm:aext}, $Z_1 \approx_{2^{-\Omega(n)}} U_{\Omega(n)}$ and is independent of $X_0$. Note that $H(\Enc(L_0(X_0))) = H(L_0(X_0)) > \gamma n/2$. Thus by Theorem~\ref{thm:samp}, with probability $1-2^{-\Omega(n)}$ over the fixing of $Z_1$, we have that $\samp(\Enc(L_0(X_0)), Z_1)$ is an affine source with entropy $\gamma n/4$. We can now fix $Z_1, X_1$. Note that conditioned on the fixing of $Z_1$, $\samp(\Enc(L_0(X_0)), Z_1)$ is a deterministic function of $X_0$, thus further fixing $X_1$ does not affect $\samp(\Enc(L_0(X_0)), Z_1)$, but this fixes $\samp(\Enc(L_1(X_1))+b, Z_1)$. Therefore, in this case we have $\widetilde{X_1} \neq \widetilde{X_1}'$ with probability at least $1-2^{-\gamma n/4}=1-2^{-\Omega(n)}$ over the further fixing of $X_0$.
\end{description}

We now condition on a particular fixing of $(X_0, X'_0, X_1, X'_1, \widetilde{X_0}, \widetilde{X_1}, \widetilde{X_0}', \widetilde{X_1}')$ such that $\widetilde{\alpha} \neq \widetilde{\alpha}'$. Note that all these are linear functions of $X$, and the total size is at most $12 \gamma n$. Thus by Lemma~\ref{lem:affindep}, we can view the remaining blocks of $X$ as independent affine sources with entropy deficiency at most $12 \gamma n$. Since $X'$ is an affine function of $X$, for any $i, j \in \N$ with $3 \leq i, j \leq 14$, we use $L^{ij}$ to denote the affine function that corresponds to the contribution of $X^j$ in $X_i'$, and use $\hat{L}^i$ to denote the affine function that corresponds to the contribution of $\hat{X}$ in $X_i'$. Thus we have for any $i$,

\[X'_i =\sum_{3 \leq j \leq 14}L^{ij}(X_j)+\hat{L}^i(\hat{X}).\]

%We now fix $L^{(13)j}(X_j)$ for all $3 \leq j \leq 14$ with $j \neq 13$, and $\hat{L}^{13}(\hat{X})$. Notice that all these random variables are on $\alpha n$ bits, thus each of them has entropy at most $\alpha n$. Therefore, conditioned on this fixing, the blocks of $X$ are still independent affine sources, with entropy deficiency at most $12 \gamma n+\alpha n$, while $X_{13}$ has entropy at least $\alpha n- 12\gamma n$. Furthermore, now $X'_{13}$ is a deterministic function of $X_{13}$.

We now again have two cases.

\begin{description}
    \item[Case 1. ] There exists an $i$ with $3 \leq i \leq 12$, and some $j \neq i$ such that $H(L^{ij}(X_j)) \geq 13 \gamma n$, or $H(\hat{L}^i(\hat{X})) \geq 13 \gamma n$. Then we have $H(X_i)+H(X_i') \geq \alpha n-12 \gamma n+13 \gamma n \geq \alpha n+\gamma n$. Since $X'_i$ is on $\alpha n$ bits, for any $x'_i \in \bits^{\alpha n}$ we have $H(X_i | X'_i=x'_i) \geq \gamma n$. Therefore by Theorem~\ref{thm:aext} we have 
\[(V_{i-2}, V'_{i-2}) \approx_{2^{-\Omega(n)}} (U_{\Omega(n)}, V'_{i-2}).\]

\item[Case 2.]  Otherwise, we can fix all the $L^{ij}(X_j)$ with $3 \leq i \leq 12,i \neq j$ and $\hat{L}^i(\hat{X})$. Conditioned on these fixings, the blocks of $X$ are still independent affine sources, and any $X_i$ with $3 \leq i \leq 12$ has entropy at least $\alpha n-12 \gamma n-10 \cdot 13 \gamma n=\alpha n- 142 \gamma n$. Moreover, conditioned on these fixings, each $X'_i$ with $3 \leq i \leq 12$ is a deterministic function of $X_i$. By Theorem~\ref{thm:czext}, as long as $\alpha$ is large enough compared to $\gamma$, we now have 
\[(V_{11}, V'_{11}) \approx_{2^{-\Omega(n)}} (U_{\Omega(n)}, V'_{11}).  \]

\end{description}
Therefore, in summary, there exists an $i \in [11]$ such that $(V_{i}, V'_{i}) \approx_{2^{-\Omega(n)}} (U_{\Omega(n)}, V'_{i})$.

Without loss of generality assume $i=1$. Note that $(V, V')$ is a deterministic function of $\{X_i, X'_i\}_{3 \leq i \leq 12}$, and $X_{13}$ has entropy at least $30\alpha n-12 \gamma n$. Let $L$ be the affine function that corresponds to the contribution of $X_{13}$ in $\{X'_i\}_{3 \leq i \leq 12}$. Thus by Lemma~\ref{lem:affinecond}, there exist independent affine sources $A, B$ such that $X_{13}=A+B$ and $L(A)=0$. Therefore, $A$ is independent of $(B, \{X_i, X'_i\}_{3 \leq i \leq 12})$ (since the blocks of $X$ are independent), and $H(A) \geq 30\alpha n-12 \gamma n-2 \cdot 10\alpha n=10 \alpha n-12 \gamma n$.\ Further fix the linear contribution of $X_{14}$ and $\hat{X}$ in $\{X'_i\}_{3 \leq i \leq 13}$. Since both $X_{14}$ and $\hat{X}$ have large size, this fixing does not cause them to lose much entropy. Moreover, conditioned on this fixing, $X'_{13}$ is a deterministic affine function of $\{X_i\}_{3 \leq i \leq 13}$. Therefore we can write $X'_{13}=A'+B'$, where $A'$ is an affine function of $A$ and $B'$ is an affine function of $(B, \{X_i\}_{3 \leq i \leq 12})$. Thus $(A, A')$ is independent of $(B, B', \{X_i, X'_i\}_{3 \leq i \leq 12}, V, V')$. We now fix $V'_1$, and this fixing does not affect the previous property. Next, by adjusting parameters and always using a strong linear seeded extractor in the $t$-affine correlation breaker when extracting from $X_{13}$ as in \cite{ChattopadhyayL22}, and noticing $t=20$, we can fix $R'_1=\affcb(X'_{13}, V'_1, 1)$ by gradually fixing at most $\alpha n$ bits of random variables, while preserving the previous property and ensuring that $H(A) \geq 10 \alpha n-12 \gamma n-\alpha n=9 \alpha n-12 \gamma n$. Note that $V_1 \approx_{2^{-\Omega(n)}} U_{\Omega(n)}$, by Theorem~\ref{thm:acbmain}, we have
\[(R_1, \{R_i\}_{2 \leq i \leq 11}, \{R'_i\}_{2 \leq i \leq 11}) \approx_{2^{-\Omega(n)}} (U_{\Omega(n)}, \{R_i\}_{2 \leq i \leq 11}, \{R'_i\}_{2 \leq i \leq 11}).\]
Therefore we also have $(R, R') \approx_{2^{-\Omega(n)}} (U_{\Omega(n)}, R').$

Finally, notice that $(R, R')$ is a deterministic function of $\{X_i, X'_i\}_{3 \leq i \leq 13}$, and $X_{14}$ has entropy at least $100\alpha n-12\gamma n$. Thus as long as $\alpha$ is large enough, by Lemma~\ref{lem:affinecond}, the fact that $\iext$ is a strong linear seeded extractor, and using a similar argument as above, we have that 
\[(W, W') \approx_{2^{-\Omega(n)}} (U_{\Omega(n)}, W').\]
\end{proof}

\subsection{Efficiently Sampling the Pre-image}
We now show that given any output of the non-malleable affine extractor in Algorithm~\ref{alg:anmext}, one can efficiently uniformly sample from the pre-image. We have the following  lemma.

\BL \label{lem:samp}
Given any arbitrary fixing of $(\{X_i\}_{0 \leq i \leq 13}, W)$, there is an efficient procedure to uniformly sample from the pre-image $X$. Moreover, for any fixing of $(\{X_i\}_{0 \leq i \leq 13}, W)$, the pre-image has the same size.
\EL

\begin{proof}
Given $(\{X_i\}_{0 \leq i \leq 13}, W)=(\{x_i\}_{0 \leq i \leq 13}, w)$, we sample from the corresponding $(X_{14}, \hat{X})$ as follows. First we compute the corresponding $z_0, z_1$, and use them to sample from $\overline{x}_2=\Enc(x_2)$ to get $\widetilde{\alpha}=x_0 \circ x_1 \circ \widetilde{x_0} \circ \widetilde{x_1}$. Next, we compute $\{v_i\}_{i \in [11]}$, $\{r_i\}_{i \in [11]}$, and $r=\sum_{i \in [11]} r_i$. Now note that $w=\iext(x_{14}, r)$, therefore by Theorem~\ref{thm:iext} we can efficiently and uniformly sample the pre-image of $w$, which is $X_{14}$, by inverting a system of linear equations. Also, Theorem~\ref{thm:iext} guarantees that for any $(r, w)$ the pre-image has the same size. 

With $X_{14}$ sampled, we continue to sample $\hat{X}$ according to the linear constraints imposed by the linear code: $\widetilde{X_0}=\widetilde{x_0}$ and $\widetilde{X_1}=\widetilde{x_1}$. This gives us  $2s \leq 2\gamma n $ linear equations, with  $\hat{X}$ being the variables. Furthermore, the length of $\hat{X}$ is $n'=(1-4 \gamma)n$. Thus, the linear equations correspond to a $n' \times 2s$ submatrix in the generator matrix of the linear code. By Theorem~\ref{thm:bcode}, as long as $\gamma$ is small enough, the $2s$ columns are linearly independent. Hence, we can efficiently sample $\hat{X}$ by inverting the system of linear equations, and moreover for any fixing of $(\{X_i\}_{0 \leq i \leq 14}, W)=(\{x_i\}_{0 \leq i \leq 13}, w)$ the pre-image always has the same size.
\end{proof}

We now have the following theorem.

\BT \label{thm:msample}
Given any output $W=w$ of the non-malleable affine extractor, there is an efficient procedure to uniformly sample from the pre-image.
\ET

\begin{proof}
The sampling procedure is as follows. We first uniformly randomly generate $(\{X_i\}_{0 \leq i \leq 13}, W)$, then we use Lemma~\ref{lem:samp} to generate $X$. By Lemma~\ref{lem:samp}, for any fixing of $(\{X_i\}_{0 \leq i \leq 13}, W)$, the pre-image has the same size. Thus this procedure indeed uniformly samples from the pre-image $X$ of $W=w$.
\end{proof}
\section{Non-Malleable Somewhere Condenser}\label{sec:nmswcond}
In this section we present out non-malleable somewhere condenser.

\BD [non-malleable somewhere condenser with advice] A function

\[\advsrcond: \bits^n  \times \bits^{a} \to (\bits^m)^t\] is called a $(k, k', \eps)$ non-malleable somewhere condenser with advice if the following holds. Let $X, X'$ be $n$-bit
random variables such that $H_{\infty}(X) \geq k$. Then, for any pair of distinct $a$-bit strings $\alpha, \alpha'$, we have that
$(\advsrcond(X, \alpha), \advsrcond(X', \alpha'))$ is $\eps$-close to a convex combination of random variables $(Z^i, Z'^i) \in (\bits^m)^t \times (\bits^m)^t$ such that for any $i$, there exists $j \in [t]$ so that for any $z \in \supp(Z'^i_j)$, we have $Z^i_j | (Z'^i_j=z)$ is a $k'$-source. 
\ED

We have the following lemma.

\BL \label{lemma:smcond}
Suppose for some constants $\ell \in \N, \gamma >0$ there is an explicit construction of an $\ell$-source non-malleable extractor for min-entropy $(1-\gamma)n$ , with output length $m=\Omega(n)$ and error $\eps=2^{-\Omega(n)}$, then there is a constant $\beta>0$ and an explicit construction of a $((1-\beta)n, \beta m, 2^{-\Omega(n)})$ non-malleable somewhere condenser with advice $\advsrcond: \bits^n \times \bits^{\beta n} \to (\bits^m)^{\ell+1}$ with $m=n/\ell$.
\EL

Let $\nmlext$ be the $\ell$-source non-malleable extractor. Our construction of the non-malleable somewhere condenser with advice is simple, as follows.

\begin{algorithm}[H]
    \caption{$\advsrcond(x, \alpha)$}
    \label{alg:advsrcond}
    \begin{algorithmic}
        \medskip
        \State \textbf{Input:} $x \in \bits^n$ --- an $n$ bit string; $\alpha \in \bits^a$, a given advice string; $\ell \in \N$, a given parameter.
        \State \textbf{Output:} $z \in (\bits^m)^{\ell+1}$ --- a matrix with $\ell+1$ bit strings of length $m$, where $m = n/\ell$.
        \\\hrulefill 
        \State \textbf{Sub-Routines and Parameters: } \\
        Let $\nmlext$ be an $\ell$-source non-malleable extractor. 
        \\\hrulefill \\
        
         \begin{enumerate}
        \item Divide $x$ evenly into $\ell$ blocks $x=x_1 \circ \cdots \circ x_{\ell}$, where each block has $m=n/\ell$ bits. 
        \item For any $i \in [\ell]$, let $z_i=x_i$. 
        \item Let $z_{\ell+1}=\nmlext(x_1 \circ \alpha, \cdots, x_{\ell} \circ \alpha)$, padding $0$'s to make the length $n/\ell$ if necessary.
        \end{enumerate}

    \end{algorithmic}
\end{algorithm}

\begin{proof}[Proof of Lemma~\ref{lemma:smcond}]
We show the function given above is such a non-malleable somewhere condenser with advice.

    Given an $(n, (1-\beta) n)$ source $X$ with $\beta>0$, and $X=X_1 \circ X_2 \circ \cdots \circ X_{\ell}$, without loss of generality we can assume that $X$ is the uniform distribution over a set $S \subseteq \zo^n$ with $|S|=2^{(1-\beta) n}$, and $X'$ is a deterministic function of $X$ (we can fix any additional randomness), i.e., $X'=f(X)$. Consider $X'=X'_1 \circ X'_2 \circ \cdots \circ X'_{\ell}$. For $i \in [\ell]$ define $H_i = \{(y, y') \in \zo^{2m}: \Pr[(X_i, X'_i) = (y, y')] \geq 2^{-(1+3\beta) m} \}$, which corresponds to the heavy elements in $(X_i, X'_i)$. Notice that this implies for every $i$, $|H_i| \leq 2^{(1+3\beta) m}$. Let $\tau=2^{-\beta m}$. We define the following sets.

    \begin{enumerate}
        \item $S' = \{x \in S : \exists i, (x_i, x'_i) \notin H_i\}$. 
        \item For any $x \in S'$, define $I(x)$ to be the smallest $i$ such that $(x_i, x'_i) \notin H_i$, and $T_i=\{x \in S': I(x)=i\}$. Let $B=\{i \in [\ell]: |T_i| < 2^{(1-\beta) n-\beta m} \}$, and define $\widetilde{S}=S' \setminus (\cup_{i \in B} T_i )$. Note that $|\cup_{i \in B} T_i| \leq \ell \tau |S|$.
        \item $S'' = \{x \in S : \forall i, (x_i, x'_i) \in H_i\}= S \setminus S'$.
    \end{enumerate}
    
    Note that for any $x \in \widetilde{S}$, we have $I(x) \notin B$. Let $\widetilde{X}$ be the uniform distribution over $\widetilde{S}$, and $\widetilde{X}'=f(\widetilde{X})$. For any $i \in [\ell] \setminus B$, conditioned on $I(\widetilde{X})=i$, i.e., $\widetilde{X} \in T_i$, we have that for any $(x_i, x'_i) \in \supp(\widetilde{X}_i,\widetilde{X}'_i)$,
    \begin{align*}
        \Pr[(\widetilde{X}_i,\widetilde{X}'_i)=(x_i, x'_i)] & = \Pr[(X_i, X'_i)=(x_i, x'_i) | X \in T_i] \leq \frac{\Pr[(X_i, X'_i) = (x_i, x'_i)]}{\Pr[X \in T_i]} \\ & \leq \frac{2^{-(1+3\beta) m}}{2^{-\beta m}} = 2^{-(1+2\beta) m}.
    \end{align*}
    
    Thus $(\widetilde{X}_i, \widetilde{X}'_i)$ has min-entropy at least $(1+2 \beta) m$. By Lemma~\ref{lem:condition}, with probability at least $1-2^{-\beta m}$ over the fixing of $\widetilde{X}'_i$, the min-entropy of $\widetilde{X}_i$ is at least $(1+2 \beta) m-m-\beta m=\beta m$. %Hence $\widetilde{X}$ is an elementary somewhere-$(1+\lambda/2)\delta m$ source.
    
    We now have two cases.

    \paragraph{Case 1.} $\Pr[X \in S'] \geq 1-\tau$. In this case, notice that $\widetilde{X}$ is $\ell \tau+\tau=(\ell+1)\tau$-close to $X$. Further conditioning on the events of $I(\widetilde{X})=i$ and that the min-entropy of $\widetilde{X}_i$ given $\widetilde{X}'_i$ is at least $\beta m$, we see that $\advsrcond(X)$ satisfies the conditions of the non-malleable somewhere condenser with error $(\ell+2)\tau=2^{-\Omega(n)}$.

\paragraph{Case 2.} $\Pr[X \in S''] \geq \tau$. In this case, notice that $|S''| \geq \tau |S| = 2^{(1-\beta)n- \beta m}$. %Also, $S''$ is a subset of $H_1 \times H_2 \times \cdots \times H_{\ell}$, so 

For each $i \in [\ell]$, define the following set 
$$V_i=\{y \in \bits^m: \exists > 2^{\beta n+ 6\beta m} \text{ strings } y' \in \bits^m \text{ such that } y \circ y' \in H_i \}.$$

Note that this implies $|V_i| < |H_i|/2^{\beta n+ 6\beta m} \leq 2^{(1-3\beta)m-\beta n}$. Define the set 
$$V=\{x \in S'': \exists i \text{ such that } x_i \in V_i\},$$ 

and notice $|V| < \ell 2^{(1-3\beta)m-\beta n}2^{(\ell-1)m}=\ell 2^{(1-\beta)n- 3\beta m} < 2^{(1-\beta)n-2\beta m}.$ Hence $|V|/|S''| < 2^{-\beta m}$. Let $S^*=S'' \setminus V$ and $X^*$ be the uniform distribution over $S^*$, let $X'^*=f(X^*)$. Then

$$|S^*| \geq (1-2^{-\beta m})2^{(1-\beta)n- \beta m} > 2^{n-2\ell \beta m}.$$
For any $i \in [\ell]$, let $S_i$ be the support of $X^*_i$. Notice that $S^*$ is a subset of $\Pi_{i \in [\ell]} S_i$, thus we have

\[\Pi_{i \in [\ell]} |S_i|=|S_1 \times S_2 \times \cdots \times S_{\ell}| \geq |S^*|.\]

Hence for any $i \in [\ell]$, %conditioned on $X \in S^*$. Notice that we have for any $i \in [\ell]$,

$$|S_i| \geq |S^*|/2^{(\ell-1)m} >2^{n-2\ell \beta m}/2^{(\ell-1)m}>2^{(1-2 \ell \beta) m}.$$

On the other hand, notice that
$$|S_1 \times S_2 \times \cdots \times S_{\ell}| \leq 2^{\ell m} \leq 2^{2 \ell \beta m}|S^*|.$$

%$$|S_1 \times S_2 \times \cdots \times S_{\ell}| \leq |H_1 \times H_2 \times \cdots \times H_{\ell}| \leq 2^{\ell(1+3\beta) m} \leq 2^{5 \ell \beta m}|S^*|.$$

We now have the following claim.

\BCM
There exists a finite set $\Q$, a family of sets $\{V_{q} \subseteq \bits^n, q \in \Q\}$, and a family of functions $\{g_q: \bits^n \to \bits^n, q \in \Q\}$ such that the following holds:
\begin{itemize}
   \item For any $q \in \Q$, $g_q =(g^1_q \circ g^2_q \circ \cdots \circ g^{\ell}_q)$, where each $g^i_q$ is a deterministic function from $\bits^m$ to $\bits^m$.
    \item For any $q \in \Q$, $|V_q| \geq 2^{-10 \ell \beta n} \Pi_{i \in [\ell]}|S_i|$.
    \item For any $q \in \Q$, $(V_q, g_q(V_q)) \subseteq \supp(X^*, X'^*)$. Furthermore, for any $q_1 \neq q_2 \in \Q$, $(V_{q_1}, g_{q_1}(V_{q_1})) \cap (V_{q_2}, g_{q_2}(V_{q_2})) = \emptyset$.
    \item $|(S^*, f(S^*)) \setminus \cup_{q \in \Q}(V_q, g_q(V_q))| \leq 2^{-\beta n} |S^*|.$
\end{itemize}
\ECM

%Since $X'=f(X)$, we also have $X'^*=f(X^*)$.
\begin{proof}[Proof of the claim.]
For any $i \in [\ell]$ and any $y \in S_i$, define $W^y_i=\{y' \in \bits^m: y \circ y' \in H_i\}$. By definition we have that for any $i \in [\ell]$ and any $y \in S_i$, $|W^y_i| \leq 2^{\beta n+ 6\beta m}$. We construct the sets $\{V_q, q \in \Q\}$ and the functions $\{g_q, q \in \Q\}$ as follows.

Initially set $\hat{S}=(S^*, f(S^*))=\supp(X^*, X'^*)$. As long as $|\hat{S}| > 2^{-\beta n} |S^*|$, consider a random function $g=(g^1 \circ g^2 \circ \cdots \circ g^{\ell})$ where for any $i \in [\ell]$ and any $y \in S_i$, let $g^i(y)$ be a random element independently uniformly chosen from $W^y_i$. For all other $y \in \bits^m$ let $g^i(y)=0^m$. Notice now we have that for any $x \in S^*$ and any $i \in [\ell]$, $x_i \in S_i$ and $(x_i, x'_i) \in H_i$. Thus for any $x \in S^*$ we have

$$\Pr[(x, x')=(x, g(x))] \geq (2^{-\ell(\beta n+ 6\beta m)}) \geq 2^{-7 \ell \beta n}.$$

Thus, by linearity of expectation, there exists a function $g$ and a set $V \subseteq \bits^n$ with $|V| \geq 2^{-7 \ell \beta n}|\hat{S}| \geq 2^{-8 \ell \beta n}|S^*| \geq 2^{-10 \ell \beta n} \Pi_{i \in [\ell]}|S_i|$ such that $(V, g(V)) \subseteq \hat{S} \subseteq \supp(X^*, X'^*)$.\ Add this function and the set $V$ to the family $\{g_q\}$ and $\{V_q\}$, let $\hat{S} \leftarrow \hat{S} \setminus (V, g(V))$ and repeat the process.

It is easy to see that the process terminates in a finite number of steps, and the sets $(V_q, g_q(V_q))$ are disjoint. When the process terminates, the final set $\hat{S}=(S^*, f(S^*)) \setminus \cup_{q \in \Q}(V_q, g_q(V_q))$ has size at most $2^{-\beta n} |S^*|$.
\end{proof}

We now consider the sources $(Y_1, Y_2, \cdots, Y_{\ell})$ where each $Y_i$ is the independent uniform distribution over $S_i$. Notice that the entropy rate of each $(Y_i \circ \alpha)$ is at least $\frac{(1-2 \ell \beta) m}{m+\beta n} \geq (1-3 \ell \beta)$. By our assumption of the $\ell$-source non-malleable extractor, as long as $3 \ell \beta \leq \gamma$, for any $q \in \Q$, 
\begin{align*}
& (\nmlext(Y_1 \circ \alpha, Y_2 \circ \alpha, \cdots, Y_{\ell} \circ \alpha), \nmlext(g_q(Y_1) \circ \alpha', g_q(Y_2) \circ \alpha', \cdots, g_q(Y_{\ell}) \circ \alpha')) \\ \approx_{\eps'} & (U_{m'}, \nmlext(g_q(Y_1) \circ \alpha', g_q(Y_2) \circ \alpha', \cdots, g_q(Y_{\ell}) \circ \alpha')),    
\end{align*}
for some $m'=\Omega(m)$ and $\eps'=2^{-\Omega(m)}$. Now for any $q \in \Q$, let $X_q$ be the uniform distribution over $V_q$, and $X'_q=f(X_q)=g_q(X_q)$. Since $|V_q| \geq 2^{-10 \ell \beta n} \Pi_{i \in [\ell]}|S_i|$ for any $q \in \Q$, by Lemma~\ref{lem:condition1}, for any $\e>0$ we have that $((\advsrcond(X_q, \alpha), (\advsrcond(X'_q, \alpha'))$ is $\e+2^{10 \ell \beta n}\eps'$-close to a distribution $(Z_q, Z'_q)$ such that for any $z \in \supp(Z'_q)$, $Z_q|(Z'_q=z)$ has min-entropy $m'-10 \ell \beta n-\log(1/\e)$. Thus, by taking $\e=2^{-\Omega(n)}$ to be large enough and $\beta$ to be a small enough constant, $((\advsrcond(X_q, \alpha), (\advsrcond(X'_q, \alpha'))$ is $2^{-\Omega(n)}$-close to a distribution $(Z_q, Z'_q)$ such that for any $z \in \supp(Z'_q)$, $Z_q|(Z'_q=z)$ has min-entropy $\beta m$. If $m' < m$, then we pad $0$'s at the end to increase the length to $m$ without affecting the property of conditional entropy.

Finally, notice that $X$ is $\ell \tau+2^{-\beta m}+2^{-\beta n}=2^{-\Omega(n)}$-close to a convex combination of $\widetilde{X}$ and $\{X_q, q \in \Q\}$. Thus the lemma also holds in this case.
\end{proof}
%After getting the H's, first remove heavy elements in the first part for each $H_i$, and argue that the number of such elements are small even compared to $S_i$. Then argue the subsource still has high entropy, since every element now has pr $\leq $, but the total probability mass $\geq (|S_i|-removed elements) \cdot 2^{-|H_i|}$. So we can apply the nm extractor. Finally, consider two parts of $S$, the elements in the removed parts only has small probability mass (crucially use the fact that in $S$ the mapping from first part to send is one to one). The elements in the remained subsource has large probability mass since the rest of $S$ has size $\geq |S|-removed elements$ and use a union bound for the removed elements, while the subsource's size dereases. Need to choose a small output size of the nm ext for conditioning.

%Next, we use the following theorem proved by Chattopadhyay and Zuckerman \cite{CZ14}.

%\BT [\cite{CZ14}]
%There is a constant $0<\gamma<1$ and an explicit non-malleable $10$-source extractor $\czext$ for $(n, (1-\gamma)n)$ sources with error $2^{-\Omega(n)}$ and output length $\Omega(n)$.
%\ET

Combined with Theorem~\ref{thm:czext}, this immediately gives the following theorem.

\BT \label{thm:smnmcond}There is a constant $\beta>0$ and an explicit construction of a $((1-\beta)n, \beta m, 2^{-\Omega(n)})$ non-malleable somewhere condenser with advice $\advsrcond: \bits^n \times \bits^{\beta n} \to (\bits^m)^{11}$ with $m=n/10$.
\ET
\section{Non-Malleable Correlation Breaker with Advice}\label{sec:advcb}
With the previous construction of a non-malleable somewhere condenser, we can now construct a non-malleable correlation breaker with advice (Definition~\ref{def:advcb}). 

%We now use the non-malleable independence preserving merger to construct an improved correlation breaker with advice. A correlation breaker, as its name suggests, uses independent randomness to break the correlations between several correlated random variables. A prototype correlation breaker was first constructed implicitly in the author's work \cite{Li13b}, and then later strengthened and formally defined in \cite{Cohen15}. A correlation breaker with advice additionally uses some string as an advice. This object was first introduced and used without its name in \cite{CGL15}, and then explicitly defined in \cite{Coh15nm}. We have the following definition.

We construct a correlation breaker with advice such that $X, X', Y, Y'$ are all $d$-bit random variables with $H_{\infty}(X) \geq 0.9d$ and $H_{\infty}(Y) \geq 0.9d$. The construction is given below. %For simplicity, when we say a strong seeded extractor for min-entropy $k$, we mean a strong average case seeded extractor for average conditional min-entropy $k$.  

\begin{algorithm}[H]
    \caption{$\acb(x)$}
    \label{alg:acb}
    \begin{algorithmic}
        \medskip
        \State \textbf{Input:} $x, y \in \bits^{d}$ --- two $d$ bit strings; $\alpha \in \bits^a$, a given advice string.
        \State \textbf{Output:} $z \in \bits^m$ --- a string with length $m=\Omega(d)$.
        \\\hrulefill 
        \State \textbf{Sub-Routines and Parameters: } \\
       \item Let $\bip$ be the two source extractor from Theorem~\ref{thm:ip}, set up to extract from two $0.35d$-bit sources and output $0.1d$ bits.  
       %\item Let $\Ext$ be a strong seeded extractor from Theorem~\ref{thm:optext}, which uses $0.05 d$ random bits to extract from a $(d, 0.2d)$ source and outputs $0.05d$ bits with error $2^{-\Omega(d)}$.
       \item Let $\advsrcond$ be the non-malleable somewhere condenser with advice from Theorem~\ref{thm:smnmcond}. 
       \item Let $\Raz$ be the two source extractor from Theorem~\ref{thm:Razext}.
       \item Let $\zuc$ be the somewhere condenser from Theorem~\ref{thm:swcondenser}, which converts a weak source with entropy rate $\beta$ to a somewhere rate $0.8$ source, where $\beta$ is the constant in Theorem~\ref{thm:smnmcond}.
       \item Let $\affcb$ be the $t$-affine correlation breaker with advice from Theorem~\ref{thm:acbmain}, for some constant $t$ to be chosen later. 
        \\\hrulefill \\
        
        %Set $x^0=x$ and let $i=0$. Initially $x^i$ has only $n_0 = 1$ row. 
         \begin{enumerate}
        \item Let $x_1$ be a slice of $x$ with length $0.35d$, and $y_1$ be a slice of $y$ with length $0.35d$. Compute $v=\bip(x_1, y_1)$.
        %\item Compute $w=\Ext(x, v)$.
        \item Compute $r=(r_1, \cdots, r_{\ell})=\advsrcond(v, \alpha)$ where $\ell=11$. 
        \item For each $i \in [\ell]$, compute $\zuc(r_i)$ which outputs $D=O(1)$ rows with length $\Omega(d)$. Let $s$ be the concatenation of all the rows from all the outputs, that is, $s$ consists of $D\ell$ rows.
        \item For each $j \in [D \ell]$, compute $w_j=\Raz(y, s_j)$ and output $m'=\Omega(d) \leq 0.01d$ bits. %Let $h$ be the concatenation of all $h_j$'s.
        \item For each $j \in [D \ell]$, compute $z_j=\affcb(x, w_j, j)$ with $t=2(D\ell-1)$ and output $m=\Omega(d) \leq 0.1 d$ bits. Finally output $z=\oplus_j z_j$.
        \end{enumerate}

    \end{algorithmic}
\end{algorithm}

We now have the following lemma.

\BL \label{lem:advcb}
There exists a constant $C>1$ such that for any $0< \e< 1/2$ and any $a, d \in \N$ with $d \geq C(a+\log(1/\e))$, there is an explicit construction of a $(0.9d, 0.9d, \e)$ strong correlation breaker with advice $\acb: \bits^d \times \bits^d \times \bits^a \to \bits^{\Omega(d)}$.
\EL

\begin{proof}
Consider $X, X'$ and $Y, Y'$ as in the definition of the correlation breaker. Note that the slice $X_1$ and $Y_1$ each has min-entropy at least $0.25d$. Thus by Theorem~\ref{thm:ip}, we have 
\[(V, Y_1) \approx_{2^{-\Omega(d)}} (U_{0.1 d}, Y_1).\]

We now fix $(Y_1, Y'_1)$. Conditioned on this fixing, $(X, X')$ is still independent of $(Y, Y')$, and now $(V, V')$ are deterministic functions of $(X, X')$ respectively, thus they are independent of $(Y, Y')$. Moreover, with probability $1-2^{-\Omega(d)}$ over this fixing, we have $V \approx_{2^{-\Omega(d)}} U_{0.1 d}$, and further by Lemma~\ref{lem:condition}, $Y$ has min-entropy at least $0.9d -2 \cdot 0.35 d-0.05d =0.15d$. 

%By Theorem~\ref{thm:optext}, we now have

%\[(W, V) \approx_{2^{-\Omega(d)}} (U_{0.05 d}, V).\]

%Next we fix $(V, V')$. Conditioned on this fixing, $(X, X')$ is still independent of $(Y, Y')$, and now $(W, W')$ are deterministic functions of $(X, X')$ respectively, thus they are independent of $(Y, Y')$. Moreover, with probability $1-2^{-\Omega(d)}$ over this fixing, we have $W \approx_{2^{-\Omega(d)}} U_{0.1 d}$, and further by Lemma~\ref{lem:condition}, $Y$ has min-entropy at least $0.9d -2 \cdot 0.05 d-0.1d =0.7d$.

We proceed as if $V$ is uniform, since this only adds $2^{-\Omega(d)}$ to the final error. Now by Theorem~\ref{thm:smnmcond}, as long as $a \leq \beta (0.1 d)$ where $\beta$ is the constant in Theorem~\ref{thm:smnmcond}, we have that $(R=\advsrcond(V, \alpha), R'=\advsrcond(V', \alpha'))$ is $2^{-\Omega(d)}$-close to a convex combination of random variables $(R^i, R'^i) \in (\bits^{0.01d})^{11} \times (\bits^{0.01d})^{11}$ such that for any $i$, there exists $j \in [11]$ so that for any $r^i_j \in \supp(R'^i_j)$, we have $R^i_j | (R'^i_j=r^i_j)$ has min-entropy at least $0.01 \beta d$. We now ignore the error and slightly abuse  notation by treating $(R, R')$ to have this property, since this only adds $2^{-\Omega(d)}$ to the final error.

Without loss of generality assume $j=1$, i.e., for any $r_1 \in \supp(R'_1)$, we have $R_1 | (R'_1=r_1)$ has min-entropy at least $0.01 \beta d$. We first fix $R'_1$. Conditioned on this fixing, $R_1$ still has min-entropy at least $0.01 \beta d$. By Theorem~\ref{thm:swcondenser}, one of the rows in $S$, without loss of generality assume $S_1$, has entropy rate $0.8$. Since $R'_1$ is fixed, $S'_1$ is also fixed. Next we fix $W'_1=\Raz(Y', S'_1)$. Notice that at this point $W'_1$ is a deterministic function of $Y'$, thus conditioned on this fixing, $(X, X')$ is still independent of $(Y, Y')$. Furthermore, by Lemma~\ref{lem:condition}, with probability $1-2^{-0.04d}$ over this fixing, $Y$ has min-entropy at least $0.15d-0.04d-0.01d =0.1d$. Therefore by Theorem~\ref{thm:Razext}, we have 

\[(W_1, S_1) \approx_{2^{-\Omega(d)}} (U_{m'}, S_1).\]

Now fix $S_1$, and conditioned on this fixing, $(X, X')$ is still independent of $(Y, Y')$; moreover $W_1$ is now a deterministic function of $Y$, thus independent of $(X, X')$. We can now further fix $(V, V')$. Since these are deterministic functions of $(X, X')$, fixing them does not affect the above property. At the same time, by Lemma~\ref{lem:condition}, with probability $1-2^{-\Omega(d)}$ over this fixing, $X$ still has min-entropy at least $0.9d - 2 \cdot 0.1d -0.1d =0.6d$.

Ignoring all the errors for now, we have that conditioned on all these fixings, $W_1=U_{m'}$ with $m'=\Omega(d)$, $W'_1$ is fixed, and all the other $\{W_j, W'_j\}_{j \neq 1}$ are deterministic functions of $(Y, Y')$. We now fix $Z'_1$. Notice it is now a deterministic function of $X'$, therefore conditioned on this fixing, $(X, X')$ is still independent of $(Y, Y')$; moreover, by Lemma~\ref{lem:condition}, with probability $1-2^{-\Omega(d)}$ over this fixing, $X$ still has min-entropy at least $0.6d - 0.1d -0.1d =0.4d$. Finally, notice that the degree $t$ in the affine correlation breaker we need is $t=2(D\ell-1)=O(1)$, and the advice length there is $\log (D\ell)=O(1)$. Thus by Theorem~\ref{thm:acbmain}, and noticing that independent sources are a special case of the sources that satisfy Definition~\ref{def:affacb}, we have

$$(Z_1, \{Z_j, Z'_j\}_{j \neq 1}, W_1, \{W_j, W'_j\}_{j \neq 1}) \approx_{2^{-\Omega(d)}} (U_{m}, \{Z_j, Z'_j\}_{j \neq 1}, W_1, \{W_j, W'_j\}_{j \neq 1}).$$

Since conditioned on all the $\{W_j, W'_j\}_{j \in [D\ell]}$ ($W'_1$ is already fixed), $\{Z_j, Z'_j\}_{j \in [D\ell]}$ are deterministic functions of $(X, X')$, and $Z'_1$ is already fixed, by adding back all the errors we also have

$$(Z, Z', Y, Y') \approx_{2^{-\Omega(d)}} (U_{m}, Z', Y, Y').$$
 Thus we only need $\e \geq 2^{-\Omega(d)}$ and $a \leq \beta (0.1 d)$, which holds as long as $d \geq C(a+\log(1/\e))$ for some constant $C>1$.
\end{proof}

We also have the following theorem.

\BT \label{thm:advcb}
There exists a constant $C>1$ such that for any $0< \e< 1/2$ and any $a, d \in \N$ with $d \geq C(a+\log(n/\e))$, there is an explicit construction of a $(d, d, \e)$ strong correlation breaker with advice $\acb: \bits^n \times \bits^d \times \bits^a \to \bits^{\Omega(d)}$.
\ET

\begin{proof}[Sketch.]
    Given an $(n, k)$ source and a uniform random seed, we first take a small slice from the seed and convert the source into an almost uniform random string using an optimal strong seeded extractor (e.g., the one from Theorem~\ref{thm:optext}). Now conditioned on the fixing of the small slice and the slice of the tampered seed, we have two independent sources, both with high min-entropy rate. Applying Lemma~\ref{lem:advcb} now gives the theorem.
\end{proof}

\section{Two Source Non-Malleable Extractor}\label{sec:tnmext}
Here we construct our two-source non-malleable extractors. First we recall the definition below.

\begin{definition}[Two-Source Non-Malleable Extractor]\label{def:t2}
A function $\nmExt : \{ 0,1\}^{n} \times \{ 0,1\}^{n} \rightarrow \{ 0,1\}^m$ is a $(k_1, k_2, \e)$ two-source non-malleable extractor, if the following holds: Let $X, Y$ be two independent sources on $n$ bits with min-entropy $k_1$ and $k_2$ respectively, and $f, g : \zo^n \to \zo^n$ be two arbitrary tampering functions such that either $f$ or $g$ has no fixed points, then  $$ |\nmExt(X, Y) \circ \nmExt(f(X), g(Y)) - U_m \circ \nmExt(f(X), g(Y))| \leq \epsilon.$$ If $k_1=k_2=k$ then we say the extractor is a $(k, \e)$ two-source non-malleable extractor. 
\end{definition}

\subsection{The Extractor Construction}

The two-source non-malleable extractor is roughly the same as the construction in \cite{Li17}, except that we replace the correlation breaker there with our new construction from Lemma~\ref{lem:advcb}, and use the new code in Theorem~\ref{thm:bcode} for generating the advice. %For completeness, we include the algorithm below.

\begin{algorithm}[H]
    \caption{$\nmExt(x, y)$}
    \label{alg:nmtext}
    \begin{algorithmic}
        \medskip
        \State \textbf{Input:} $x, y \in \bits^{n}$ --- two $n$ bit strings.
        \State \textbf{Output:} $w \in \bits^m$ --- a bit string with length $m=\Omega(n)$.
        \\\hrulefill 
        \State \textbf{Sub-Routines and Parameters: } \\
        \item Let $0<\alpha<\eta<1/100$ be two constants to be chosen later.
       \item Let $\bip$ be the two source extractor from Theorem~\ref{thm:ip}. 
       \item Let $\acb$ be the correlation breaker with advice from Lemma~\ref{lem:advcb}.  %with error $\e=2^{-\Omega(n/\log n)}$.
       \item Let $\iext$ be the invertible linear seeded extractor form Theorem~\ref{thm:iext}.
       \item Let $\Enc$ be the encoding function of the linear code from Theorem~\ref{thm:bcode}.
       \item Let $\samp$ be the average sampler from Theorem~\ref{thm:sampler}.
       
        \\\hrulefill \\
        
        %Set $x^0=x$ and let $i=0$. Initially $x^i$ has only $n_0 = 1$ row. 
         \begin{enumerate}
        \item Divide $x$ into $x=(x_1, x_2)$ such that $x_1$ has $n_1=\alpha n$ bits and $x_2$ has $n_2=(1-\alpha)n$ bits. Similarly divide $y$ into $y=(y_1, y_2)$ such that $y_1$ has $n_1$ bits and $y_2$ has $n_2=(1-\alpha)n$ bits.

\item Compute $z=\bip(x_1, y_1)$ which outputs $r=\Omega(n) \leq \alpha n/2$ bits.

\item Let $\overline{x}_2=\Enc(x_2)$ and $\overline{y}_2=\Enc(y_2)$.
%\item Let $\F$ be the finite field $\F_{2^{\log n}}$. Let $n_0 = \frac{n_2}{\log n}$. Let $\RS: \F^{n_0} \rightarrow \F^{n}$ be the Reed-Solomon code encoding $n_0$ symbols of $\F$ to $n$ symbols in  $\F$ (we  slightly abuse the use of $\RS$ to denote both the code and the encoder). Thus $\RS$ is a $[n,n_0,n-n_0+1]_{n}$ error correcting code. Let $x'_2$ be $x_2$ written backwards, and similarly $y'_2$ be $y_2$ written backwards. Let $\overline{x}_2=\RS(x'_2)$ and $\overline{y}_2=\RS(y'_2)$.

\item Use $z$ and $\samp$ to sample $s=\Omega(n) \leq \alpha n/2$ distinct bits from $\overline{x}_2$, let the resulted string be $\widetilde{x}_2$. Similarly, use $z$ to sample $s$ distinct bits from $\overline{y}_2$ and obtain a binary string $\widetilde{y}_2$.

\item Let $\widetilde{\alpha}=x_1 \circ y_1 \circ \widetilde{x}_2 \circ \widetilde{y}_2$. Divide $x_2$ into $x_2=(x_3, x_4, x_5)$ such that $x_3$ has $n_3=\eta n$ bits, $x_4$ has $n_4=30\eta n$ bits and $x_5$ has $n_5=(1-\alpha-31\eta)n$ bits. Similarly divide $y_2=(y_3, y_4, y_5)$ such that $y_3$ has $n_3$ bits, $y_4$ has $n_4$ bits and $y_5$ has $n_5$ bits.

\item Compute $v=\acb(x_3, y_3, \widetilde{\alpha})$ which outputs $d=\Omega(n_3)=\Omega(n) \leq \eta n/10$ bits.

\item Finally compute $w=\iext(y_4, v)$ which outputs $\Omega(d) < d/2$ bits.
        \end{enumerate}

    \end{algorithmic}
\end{algorithm}

We now have the following theorem.

\BT \label{thm:nmtext}
There exists a constant $0< \gamma< 1$ such that for any $n \in \N$, there exists an explicit construction of a $((1-\gamma)n, 2^{-\Omega(n)})$ two-source non-malleable extractor with output length $\Omega(n)$.
\ET

\begin{proof}
We show that the above construction is such a two-source non-malleable extractor.\ As usual, we use letters with prime to denote random variables produced from $(X', Y')$. Without loss of generality we assume $X \neq X'$. The case where $Y \neq Y'$ can be handled by symmetry.

First we argue that with probability $1-2^{-\Omega(n)}$, we have that $\widetilde{\alpha} \neq \widetilde{\alpha}'$. To see this, note that if $X_1 \neq X_1'$ or $Y_1 \neq Y_1'$ then $\widetilde{\alpha} \neq \widetilde{\alpha}'$. Otherwise, since $X \neq X'$ we must have $X_2 \neq X_2'$. Thus by the property of our code from Theorem~\ref{thm:bcode}, $\overline{X}_2$ and $\overline{X}'_2$ must differ in $\Omega(n)$ bits. Also, since $X_1=X_1'$ and $Y_1=Y_1'$ we have $Z=Z'$. Now if $\alpha \geq 3\gamma$ then both $X_1$ and $Y_1$ have min-entropy rate at least $2/3$, thus by Theorem~\ref{thm:ip} we have

\[(Z, X_1) \approx_{2^{-\Omega(n)}} (U_r, X_1).\]

We can now fix $X_1$, and conditioned on this fixing, $(X, X')$ is still independent of $(Y, Y')$. Moreover $Z$ is a deterministic function of $Y$, thus independent of $X_2$. Therefore now we can use $Z$ to sample from $\overline{X}_2$. If $Z$ is uniform then by Theorem~\ref{thm:sampler} we know that 

\[\Pr[\widetilde{X}_2 \neq \widetilde{X}_2'] \geq 1-2^{-\Omega(r)}=1-2^{-\Omega(n)}.\]

Thus the total probability that $\widetilde{\alpha} \neq \widetilde{\alpha}'$ is at least $1-2^{-\Omega(n)}-2^{-\Omega(n)}=1-2^{-\Omega(n)}$.

Moreover, by choosing $\alpha< \eta/50$, we can ensure that $r \leq \alpha n/2 < \eta n/50$. Now by Lemma~\ref{lem:condition} we know that conditioned on the fixing of $(\widetilde{\alpha}, \widetilde{\alpha}')$, with probability $1-2^{-\Omega(n)}$, we have that $H_{\infty}(X_3) \geq \eta n-\gamma n-\alpha n-3r \geq 0.9 \eta n$ and similarly $H_{\infty}(Y_3) \geq 0.9 \eta n$. Moreover $(X, X')$ and $(Y, Y')$ are still independent.

Now we use Lemma~\ref{lem:advcb}. Note that the length of the advice string is $a=2\alpha n+2r \leq 3\alpha n$, and $X_3, Y_3$ each has $\eta n$ bits. Thus by choosing the error $\e=2^{-\Omega(n)}$ appropriately we can ensure that 

\[\eta n \geq C(\log a +\log (1/\e)),\]
where $C$ is the constant in  Lemma~\ref{lem:advcb}. When this condition holds, by Lemma~\ref{lem:advcb} we have that

\[(V, V', Y_3, Y_3') \approx_{\e} (U_d, V', Y_3, Y_3').\]

We now fix $(Y_3, Y_3')$, and conditioned on this fixing, $(X, X')$ is still independent of $(Y, Y')$. Note that now, $(V, V')$ is a deterministic function of $(X, X')$, and thus independent of $(Y, Y')$. Moreover the average conditional min-entropy of $Y_4$ is at least $n_4-\gamma n-\alpha n-2r-\eta n \geq n_4 -3\alpha n-\eta n$. Note that $n_4=30 \eta n$. Thus by choosing $\alpha< \eta/50$ we can ensure that (by Lemma~\ref{lem:condition}) with probability $1-2^{-\Omega(n)}$, $Y_4$ has min-entropy rate at least $0.95$.

Now we can fix $V'$ and then $W'$, and conditioned on this fixing, $(X, X')$ is still independent of $(Y, Y')$. Note that now, $V$ is still close to uniform, and independent of $Y_4$. Furthermore since the length of $W'$ is at most $d/2 \leq \eta n/20$, again by Lemma~\ref{lem:condition} we have that with probability $1-2^{-\Omega(n)}$, $Y_4$ has min-entropy rate at least $0.9$. Thus now by Theorem~\ref{thm:iext} we have that

\[(W, V) \approx_{2^{-\Omega(n)}} (U_{\Omega(n)}, V).\]

Note that conditioned on the fixing of $V$, $W$ is a deterministic function of $Y$. Since we have already fixed $(V', W')$, by adding back all the errors we get that 

\[(W, W', X, X') \approx_{2^{-\Omega(n)}} (U_{\Omega(n)}, W', X, X').\]
\end{proof}
%By Lemma~\ref{lem:nmeq}, the extractor is also a general non-malleable two-source extractor (Definition~\ref{def:gt2}.) for min-entropy $(1-\gamma/2)n$ with error $2^{-\Omega(n)}$.

Although not necessary for our applications, we can in fact reduce the entropy requirement of the above non-malleable two source extractor. Specifically, we have the following theorem.

\BT
   There exists a constant $C>1$ such that for any constant $0< \gamma< 1$, any $n \in \N$, and any $k \geq C \log n$, there exists an explicit construction of a $((2/3+\gamma)n, k, 2^{-\Omega(k)})$ non-malleable two-source extractor with output length $\Omega(k)$.
\ET

\begin{proof}[Sketch]
    Let $X$ be the $(n, (2/3+\gamma)n)$ source and $Y$ be the $(n, k)$ source. The construction is as follows. First, take a slice $X_1$ of length $n/3$ from $X$, apply the somewhere condenser $\zuc$ and output a constant number $D$ of rows s.t. one row has entropy rate at least $0.8$. Using each row as a seed and apply the extractor $\Raz$ from Theorem~\ref{thm:Razext} to $Y$, and get a constant number of outputs $\{V_i\}_{i \in [D]}$ with size $m=\Omega(k)$. For each $V_i$, take a small slice $W_i$ with size $\Omega(m)$ and use it as a seed to apply an optimal strong seeded extractor $\Ext$ from Theorem~\ref{thm:optext} to $X$, extracting $T_i$ which has $m$ bits. Thus, we now have $\{V_i\}_{i \in [D]}$ and $\{T_i\}_{i \in [D]}$. Take a larger slice $\widetilde{W}_i$ with size $\Omega(m)$ from each $V_i$, and a slice $\widetilde{T}_i$ with the same size from each $T_i$. Compute $H_i=\bip(\widetilde{W}_i, \widetilde{T}_i)$, use $H_i$ to sample $\Omega(k)$ bits from an asymptotically good binary encoding of $X$ and $Y$, and concatenate these strings with $\widetilde{W}_i \circ \widetilde{T}_i$ to get an advice string $\alpha_i$. Next, compute $R_i=\nmExt(V_i \circ \alpha_i, T_i \circ \alpha_i)$ which outputs $\Omega(k)$ bits. Finally, for each $i \in [D]$, compute $\affcb(Y, R_i, i)$ with $t=2(D-1)$ and output $\Omega(k)$ bits, then take the XOR of all these outputs. %of use a seeded extractor from Theorem~\ref{thm:optext} to  

    For the analysis, consider the tampered version $(X', Y')$. As usual, we gradually fix a sequence of random variables, while maintaining the property that $(X, X')$ is independent of $(Y, Y')$, and $(X, Y)$ each has enough min-entropy left. To do this, first note that $X_1$ has min-entropy at least $\gamma n$, and thus one row of the output of $\zuc$ has entropy rate at least $0.8$. Therefore by Theorem~\ref{thm:Razext}, some $V_i$ (without loss of generality assume $V_1$) is close to uniform. Next fix $(X_1, X'_1)$, and now $\{V_i\}_{i \in [D]}, \{V'_i\}_{i \in [D]}$ are deterministic functions of $(Y, Y')$. Moreover the average conditional min-entropy of $X$ left is at least $\gamma n$. Thus by Theorem~\ref{thm:optext}, $(T_1, W_1) \approx_{2^{-\Omega(k)}} (U_m, V_1)$. Now fix all the $\{W_i, W'_i\}_{i \in [D]}$, then $\{T_i\}_{i \in [D]}, \{T'_i\}_{i \in [D]}$ are deterministic functions of $(X, X')$. By limiting the size of each $W_i$, $V_1$ still has high min-entropy. Therefore we can take a larger slice $\widetilde{W}_1, \widetilde{T}_1$ and use $H_1=\bip(\widetilde{W}_1, \widetilde{T}_1)$ to sample the advice. This ensures $\alpha_1 \neq \alpha'_1$ with probability $1-2^{-\Omega(k)}$. Now fix all $\{\widetilde{W}_i, \widetilde{T}_i\}_{i \in [D]}, \{\widetilde{W}'_i, \widetilde{T}'_i\}_{i \in [D]}$ and the sampled bits, again by limiting their sizes, $T_1$ and $V_1$ still have high min-entropy rate. Now by Theorem~\ref{thm:nmtext}, $(R_1, R'_1, V_1, V'_1) \approx_{2^{-\Omega(k)}} (U, R'_1, V_1, V'_1)$. Further fix all $\{V_i\}_{i \in [D]}, \{V'_i\}_{i \in [D]}$, now the $\{R_i\}_{i \in [D]}, \{R'_i\}_{i \in [D]}$ are deterministic functions of $X$, and $Y$ still has enough min-entropy left (by limiting the size of each $V_i$). Now, as in the analysis of Theorem~\ref{thm:nmtext}, we can first fix $R'_1$ and the output $\affcb(Y, R'_1, 1)$ without causing $Y$ to lose much entropy, and the correlation breaker $\affcb$ from Theorem~\ref{thm:affcb} guarantees that the output from $(X, Y)$ is close to uniform given the output from $(X', Y')$. Since $D=O(1)$ we can afford to use outputs of size $\Omega(k)$ in all computations, and thus the final output is $\Omega(k)$ and the final error is $2^{-\Omega(k)}$.
\end{proof}

\begin{remark}
The above non-malleable two source extractor can also handle sources with uneven lengths, since the extractor $\Raz$ from Theorem~\ref{thm:Razext} can do so. We omit the details here.
\end{remark}
\subsection{Efficiently Sampling the Pre-image}
We now show that given any output of the non-malleable two-source extractor in Algorithm~\ref{alg:nmtext}, one can efficiently uniformly sample from the pre-image. We have the following  lemma.

\BL \label{lem:samp2}
Given any arbitrary fixing of $(X_1, \widetilde{X}_2, X_3, Y_1, \widetilde{Y}_2, Y_3, W)$, there is an efficient procedure to uniformly sample from the pre-image $(X, Y)$. Moreover, for any fixing of $(X_1, \widetilde{X}_2, X_3, Y_1, \widetilde{Y}_2, Y_3, W)$, the pre-image has the same size.
\EL

\begin{proof}
Given $(X_1, \widetilde{X}_2, X_3, Y_1, \widetilde{Y}_2, Y_3, W)=(x_1, \widetilde{x}_2, x_3, y_1, \widetilde{y}_2, y_3, w)$, we sample from the corresponding $(X_4, X_5, Y_4, Y_5)$ as follows. First we compute $z=\bip(x_1, y_1)$ and use it to sample from $\overline{x}_2=\Enc(x_2)$ and $\overline{y}_2=\Enc(y_2)$. Next, we compute $v=\acb(x_3, y_3, \widetilde{\alpha})$ where $\widetilde{\alpha}=x_1 \circ y_1 \circ \widetilde{x}_2 \circ \widetilde{y}_2$. Now note that $w=\iext(y_4, v)$, therefore by Theorem~\ref{thm:iext} we can efficiently and uniformly sample the pre-image of $w$, which is $Y_4$, by inverting a system of linear equations. Also, Theorem~\ref{thm:iext} guarantees that for any $(v, w)$ the pre-image has the same size. 

With $Y_4$ sampled, we continue to sample $(X_4, X_5, Y_5)$ according to the linear constraints imposed by the linear code: $\widetilde{Y}_2=y_2$ and $\widetilde{X}_2=x_2$. Consider the $Y$ part. Note that $\widetilde{Y}_2=y_2$ gives us $r \leq \alpha n/2< n/4 $ equations, and that $(Y_1, Y_3, Y_4)$ are fixed, with  $Y_5$ being the variables in the linear equations. Furthermore, the length of $Y_5$ is $n_5=n-\alpha n-\eta n-30 \eta n> 2n/3$ (as $\alpha < \eta < 1/100$). Thus, the linear equations correspond to a $n_5 \times r$ submatrix in the generator matrix of the linear code. By Theorem~\ref{thm:bcode}, as long as $\alpha$ is small enough, the $r$ columns must be linearly independent. Hence, we can efficiently sample $Y_5$ by inverting the system of linear equations, and moreover for any fixing of $(X_1, \widetilde{X}_2, X_3, Y_1, \widetilde{Y}_2, Y_3, W)=(x_1, \widetilde{x}_2, x_3, y_1, \widetilde{y}_2, y_3, w)$ the pre-image always has the same size.

The argument for sampling the $X$ part is exactly the same, except now $X$ has more free variables $(X_4, X_5)$ than $Y$.
\end{proof}

We now have the following theorem.

\BT \label{thm:msample2}
Given any output $W=w$ of the non-malleable two-source extractor, there is an efficient procedure to uniformly sample from the pre-image $(X, Y)$.
\ET

\begin{proof}
The sampling procedure is as follows. We first uniformly randomly generate $(X_1, \widetilde{X}_2, X_3, Y_1, \widetilde{Y}_2, Y_3)$, then we use Lemma~\ref{lem:samp2} to generate $(X, Y)$. By Lemma~\ref{lem:samp2}, for any fixing of $(X_1, \widetilde{X}_2, X_3, Y_1, \widetilde{Y}_2, Y_3, W)$, the pre-image has the same size. Thus this procedure indeed uniformly samples from the pre-image $(X, Y)$ of $W=w$.
\end{proof}

\section{Applications} \label{sec:app}
In this section we give various applications of our constructions in previous sections. 
\subsection{Extractors and Ramsey Graphs}
Two source non-malleable extractors can be conveniently converted to seeded non-malleable extractors, as shown in \cite{Li17}. Here we prove a slightly 
different version than that in \cite{Li17}. First we define seeded non-malleable extractors against multiper tampering.

\begin{definition} A function $\snmExt:\{0,1\}^n \times \{ 0,1\}^d \rightarrow \{ 0,1\}^m$ is a seeded $t$-non-malleable extractor for min-entropy $k$ and error $\epsilon$ if the following holds : If $X$ is an $(n, k)$ source and $\A_1, \cdots, \A_t : \{0,1\}^d \rightarrow \{0,1\}^d $ are $t$ arbitrary tampering functions with no fixed points, then
$$  \left |\snmExt(X,U_d) \scirc \{\snmExt(X,\A_i(U_d)), i \in [t]\} \scirc U_d- U_m \scirc  \{\snmExt(X,\A_i(U_d)), i \in [t]\} \scirc U_d \right | <\epsilon $$where $U_m$ is independent of $U_d$ and $X$.
\end{definition}

We now have the following theorem.

\BT \label{thm:nmconvert1}
Suppose there is a constant $\beta>0$ and an explicit non-malleable $2$-source extractor for $(n, (1-\beta)n)$ sources with error $2^{-\Omega(n)}$ and output length $\Omega(n)$. Then for any constant $\gamma>0$ there is a constant $C>0$ such that for any $0< \e < 1$ with $k \geq C t^3\log(d/\e)$ and $d=C t^3\log(n/\e)$, there is an explicit strong seeded $t$-non-malleable extractor for $(n, k)$ sources with seed length $d$, error $O(t \e)$ and output length $\frac{(1-\gamma)k}{t+1}$.%$$\Omega(\log(1/\e))$. 
\ET

To prove the theorem we first need the following lemma from \cite{Li17}.

\BL [\cite{Li17}]\label{lem:advcb2}
Suppose that there exists a constant $\beta>0$ and an explicit construction of a strong two-source non-malleable extractor $\nmExt: (\zo^n)^2 \to \zo^m$ for $(n, (1-2\beta)n)$ sources which outputs $\Omega(n)$ bits with error $2^{-\Omega(n)}$. Then given any $t \in \N$ there is an explicit function $\acb: (\zo^n)^2 \times \zo^{a} \to \zo^m$ with $m=\Omega(a)$ with the following property.

Let $X, Y$ be $2$ independent uniform strings on $n$ bits, and $\alpha, \alpha^1, \cdots, \alpha^t$ be $t+1$ strings on $a$ bits such that $\forall j \in [t], \alpha \neq \alpha^j$. Let $\{X^j\}_{j \in [t]}$ and $\{Y^j\}_{j \in [t]}$ be random variables on $n$ bits such that $(X, \{X^j\}_{j \in [t]})$ and $(Y, \{Y^j\}_{j \in [t]})$ are independent.  Let $Z=\acb(X, Y, \alpha)$ and $Z^j=\acb(X^j, Y^j, \alpha^j)$ for any $j \in [t]$. Then as long as $n \geq 2(t+1)^2 a/\beta$, we have that,

\[\left | (Z, \{Z^j\}_{j \in[t]}, Y, \{Y^j\}_{j \in[t]})-(U_m, \{Z^j\}_{j \in[t]}, Y, \{Y^j\}_{j \in[t]})\right | \leq t2^{-\Omega(a)}.\]
\EL

\begin{proof}[Proof of Theorem~\ref{thm:nmconvert1}] Let $X$ be an $(n, k)$ source and $Y$ be a uniform random seed. The construction of the seeded non-malleable extractor is as follows. 
\begin{itemize}
    \item Let $\Ext$ be the optimal seeded extractor from Theorem~\ref{thm:optext}.
    \item Let $\samp$ be the average sampler from Theorem~\ref{thm:samp}.
    \item Let $\Enc$ be the encoding function of the binary code in Theorem~\ref{thm:bcode}.
    \item Let $\Raz$ be the strong two source extractor from Theorem~\ref{thm:Razext}. 
\end{itemize}
\begin{enumerate}
    \item Take a small slice $Y'$ of $Y$ with length $d_1=O(\log(n/\e))$ and compute $Z=\Ext(X, Y')$ and output $s=O(t\log(d/\e))$ bits.
    \item Take a small slice $Z_1$ of $Z$ with length $d_2=O(\log(d/\e))$, and compute $Y_1=\samp(\Enc(Y), Z_1)$ with $d_2$ bits.
    \item Let the advice string be $\alpha=Z_1 \circ Y_1$. Take a larger slice $Z_2$ of $Z$ with length $d_3=O(t\log(d/\e))$, and a larger slice $Y_2$ of $Y$ of length $d_4=O(t \log (n/\e))$. Compute $W=\Raz(Z_2, Y_2)$ which outputs $d_2$ bits.
    \item Compute $\widetilde{Z}=\Ext(X, W)$ and $\widetilde{Y}=\Ext(Y, W)$, each outputs $d_5=O(t^2\log(d/\e))$ bits.
    \item Compute $V_1=\acb(\widetilde{Y}, \widetilde{Z}, \alpha)$ which outputs $\Omega(d_2)$ bits.
    \item Compute $V_2=\Ext(Y, V_1)$ and output $d_1$ bits.
    \item Compute $V=\Ext(X, V_2)$ and output $m=\frac{(1-\gamma)k}{t+1}$ bits.
\end{enumerate}

For the analysis, again we will gradually fix a sequence of random variables and maintaining that $(X, \{X^j\}_{j \in [t]})$ and $(Y, \{Y^j\}_{j \in [t]})$ are independent, and argue that $X$ and $Y$ has enough entropy. For simplicity we omit the first condition in the following argument. First note that by Theorem~\ref{thm:optext}, $(Z, Y') \approx_{\e} (U_s, Y')$. Thus we can fix all $(Y', \{Y'^j\}_{j \in [t]})$. Conditioned on this fixing, the $Z, \{Z^j\}_{j \in [t]}$ are deterministic functions of $(X, \{X^j\}_{j \in [t]})$. 

If for any $j \in [t]$ we have $Z_1 \neq Z^j_1$, then we also have $\alpha \neq \alpha^j$. Otherwise, by Theorem~\ref{thm:bcode} and Theorem~\ref{thm:sampler}, we have $\alpha \neq \alpha^j$ with probability $1-\e$. Thus by a union bound, we have $\alpha \neq \alpha^j$ for all $j \in [t]$ with probability $1-t\e$. We now proceed conditioned on the event that this happens, and fix all the $(Z_1, \{Z^j_1\}_{j \in [t]})$ and $(Y_1, \{Y^j_1\}_{j \in [t]})$. Conditioned on this fixing, $X$ and $Y$ still has enough entropy left.

By adjusting the size of $Z_2$ and $Y_2$, they both have entropy rate at least $2/3$. Thus by Theorem~\ref{thm:Razext}, we have $(W, Z_2) \approx_{\e} (U_{d_2}, Z_2)$ and $(W, Y_2) \approx_{\e} (U_{d_2}, Y_2)$. Thus, conditioned on the fixing of $(Z_2, \{Z^j_2\}_{j \in [t]})$, $W$ is close to uniform and is a deterministic function of $X$, hence by Theorem~\ref{thm:optext} we have $(\widetilde{Y}, W) \approx_{\e} (U_{d_2}, W)$. Similarly, we also have $(\widetilde{Z}, W) \approx_{\e} (U_{d_2}, W)$. Now we can fix all $(Z_2, \{Z^j_2\}_{j \in [t]})$ and $(Y_2, \{Y^j_2\}_{j \in [t]})$. Now $(\widetilde{Z}, \{\widetilde{Z}^j\}_{j \in [t]})$ and $(\widetilde{Y}, \{\widetilde{Y}^j\}_{j \in [t]})$ are deterministic functions of $(X, \{X^j\}_{j \in [t]})$ and $(Y, \{Y^j\}_{j \in [t]})$ respectively, so they are independent. Note that $\alpha$ has length $O(d_2)=O(\log(d/\e))$. Thus as long as $d_5=O(t^2\log(d/\e))$, by Lemma~\ref{lem:advcb2} we have that

\[(V_1, \{V^j_1\}_{j\in [t]}, \widetilde{Z}, \{\widetilde{Z}^j\}_{j \in [t]}) \approx_{O(t \e)} (U_{d_2}, \{V^j_1\}_{j\in [t]}, \widetilde{Z}, \{\widetilde{Z}^j\}_{j \in [t]}).\]

Fixing $\widetilde{Z}, \{\widetilde{Z}^j\}_{j \in [t]}$, we have $V_1, \{V^j_1\}_{j\in [t]}$ are deterministic functions of $(X, \{X^j\}_{j \in [t]})$. By a standard argument, and Theorem~\ref{thm:optext}, we now have 

\[(V_2, \{V^j_2\}_{j\in [t]}, V_1, \{V^j_1\}_{j\in [t]}) \approx_{\e} (U_{d_1}, \{V^j_2\}_{j\in [t]}, V_1, \{V^j_1\}_{j\in [t]}).\]

Further fix $V_1, \{V^j_1\}_{j\in [t]}$, we have $V_2, \{V^j_2\}_{j\in [t]}$ are deterministic functions of $(Y, \{Y^j\}_{j \in [t]})$. Thus again by a standard argument, and Theorem~\ref{thm:optext}, we now have 

\[(V, \{V^j\}_{j\in [t]}, V_2, \{V^j_2\}_{j\in [t]}) \approx_{\e} (U_{m}, \{V^j\}_{j\in [t]}, V_2, \{V^j_2\}_{j\in [t]}).\]

Adding back all the errors, and noticing that conditioned on the fixing of $V_2, \{V^j_2\}_{j\in [t]}$, we have $V, \{V^j\}_{j\in [t]}$ are deterministic functions of $(X, \{X^j\}_{j \in [t]})$. Thus we have

\[(V, \{V^j\}_{j\in [t]}, Y, \{Y^j\}_{j \in [t]}) \approx_{\e} (U_{m}, \{V^j\}_{j\in [t]}, Y, \{Y^j\}_{j \in [t]}).\]

The entropy requirement is that $k \geq (t+1)m +O(t d_5)$ and $d \geq O(t d_5+t d_4+d_1)$. Thus it is enough to have $k \geq C t^3\log(d/\e)$ and $d=C t^3\log(n/\e)$ for some constant $C>1$.
\end{proof}

\iffalse
\BT [\cite{Li17}] \label{thm:nmconvert1}
Suppose there is a constant $\gamma>0$ and an explicit non-malleable $2$-source extractor for $(n, (1-\gamma)n)$ sources with error $2^{-\Omega(n)}$ and output length $\Omega(n)$. Then there is a constant $C>0$ such that for any $0< \e < 1$ with $k \geq C t^2\log(d/\e)$ and $d=C t^2\log(n/\e)$, there is an explicit strong seeded $t$-non-malleable extractor for $(n, k)$ sources with seed length $d$, error $O(t \e)$ and output length $\Omega(k/t)$.%$$\Omega(\log(1/\e))$. 
\ET
\fi

Combined with Theorem~\ref{thm:nmtext}, this gives the following theorem.

\BT \label{thm:stnmext}
For any constant $\gamma>0$ there is a constant $C>0$ such that for any $0< \e < 1$ with $k \geq C t^3\log(d/\e)$ and $d=C t^3\log(n/\e)$, there is an explicit strong seeded $t$-non-malleable extractor for $(n, k)$ sources with seed length $d$, error $O(t \e)$ and output length $\frac{(1-\gamma)k}{t+1}$.
\ET

By using improved somewhere random condensers as samplers and following the framework in \cite{CZ15}, \cite{BDT16} proved the following theorem.

\BT [\cite{BDT16}] 
Suppose there is a function $f$ and an explicit strong seeded $t$-non-malleable extractor for $(n, k')$ sources with seed length and entropy requirement $d \geq f(t, \e), k' \geq f(t, \e)$, then for every constant $\e>0$ there exist constants $t=t(\e), c=c(\e)$ and an explicit two source extractor $\TExt: \zo^n \times \zo^n \to \zo$ for min-entropy $k \geq f(t, 1/n^c)$ with error $\e$.
\ET

Combined with Theorem~\ref{thm:stnmext}, we immediately get the following theorem.

\BT %\label{thm:2source}
For every constant $\e>0$ there exists a constant $c>1$ and an explicit two-source extractor $\TExt:  \zo^n \times \zo^n \to \zo$ for min-entropy $k \geq c \log n$, with error $\e$.
\ET

A standard argument then gives the following construction of Ramsey graphs.

\BCR
There exists a constant $c>1$ such that for every integer $N$ there exists a (strongly) explicit construction of a $K$-Ramsey graph on $N$ vertices with $K=\log^c N$. 
\ECR

We now define sumset sources, interleaved sources, and small space sources.

\BD[sumset source]
A source $X$ is a $(n, k, C)$-sumset source if there exist $C$ independent $(n, k)$-sources $\{X_i\}_{i\in [C]}$ such that $X = \sum_{i=1}^{C} X_i$.
\ED

\BD [interleaved source]
Let $X_1$ be a $(n, k_1)$-source, $X_2$ be a $(n, k_2)$-source independent of $X_1$ and $\sigma : [2n] \to [2n]$ be a permutation. Then $(X_1 \circ X_2)_\sigma$ is a $(n, k_1, k_2)$-interleaved source, or a $(n, k)$-interleaved source if $k_1 = k_2=k$.
\ED

\BD\cite{KRVZ06}
A space $s$ source $X$ is generated by taking  a random walk on a branching program of length $n$ and width $2^{s}$, where each edge of the branching program is labelled with a transition probability and a bit. Thus a bit of the source is generated for each step taken on the branching program, and the source $X$ is the concatenation of all the bits.
\ED

Following the work of Chattopadhyay and Li on extractors for sumset sources \cite{ChattopadhyayL16}, Chattopadhyay and Liao \cite{ChattopadhyayL22} generalized the above reductions for two-source extractors to the sum of two independent sources. Specifically, the prove the following theorem.

\BT [\cite{ChattopadhyayL22}] \label{thm:sumsetred}
There exists a constant $C_0 > 1$ such that the following holds.
Suppose there is a function $f$ and an explicit $(t, k', \e)$-affine correlation breaker for advice strings of length $a$, with seed length and entropy requirement $d \geq f(t, \e, a), k' \geq f(t, \e, a)$, then for every constant $\e>0$ there exist constants $t=t(\e), c=c(\e)$ such that if there exist $\bar{k}, C \in \N$ satisfying the following conditions:
\begin{itemize}
    \item $\bar{k} \geq f(Ct, 1/n^c, c \log n+\log C)$,
    \item $C \geq C_0 \log^2 \frac{k}{\log n}$,
\end{itemize}
then there exists an explicit extractor $\sExt: \zo^n  \to \zo$  for the sum of two independent $(n, k)$ sources with $k=O(Ct\bar{k}+\log n)$ and error $\e$.
\ET

To get the desired $(t, k, \e)$-affine correlation breaker, we combine Theorem~\ref{thm:affcb} with our new standard correlation breaker, Theorem~\ref{thm:advcb}. Thus we have

\BT  
For any $t$, there exists an explicit strong $t$-affine correlation breaker $\affcb :\bits^n \times \bits^d \times \bits^a \to \bits^m$ with error $O(t \e)$ for entropy $k=O(ta+tm+t^2\log(n/\e))$, where $d=O(ta+tm+t\log^3(t+1)\log(n/\e))$.
\ET

Combining the above theorem with Theorem~\ref{thm:sumsetred}, and noticing that for any constant $\e>0$, $t=t(\e), c=c(\e)$ are both constants, thus by choosing $m=1$, $\bar{k} \geq c \log n$ for a large enough constant $c>1$ and $C$ to be a large enough constant, we get the following theorem.

\BT \label{thm:sumsetext}
For every constant $\e>0$ there exists a constant $c>1$ and an explicit extractor $\sExt: \zo^n  \to \zo$ for the sum of two independent $(n, k)$ sources with min-entropy $k \geq c \log n$, and error $\e$.
\ET

Note that affine sources and interleaved sources are special cases of sumset sources, thus we have the following corollaries.

\BCR
For every constant $\e>0$ there exists a constant $c>1$ and an explicit affine extractor $\aext: \zo^n  \to \zo$ for entropy $k \geq c \log n$, with error $\e$.
\ECR

\BCR
For every constant $\e>0$ there exists a constant $c>1$ and an explicit extractor $\itext: \zo^{2n}  \to \zo$ for the interleaving of two independent $(n, k)$ sources with min-entropy $k \geq c \log n$, and error $\e$.
\ECR

Chattopadhyay and Liao \cite{ChattopadhyayL22} also showed an improved reduction from small space sources to sumset extractors. Specifically, they prove

\BL Every space-$s$ source $X \in \zo^n$ with min-entropy at least $k = k_1 +k_2 + 2s+ 2 \log(n/\e)$ is $3\e$-close to a convex combination of sources of the form $X_1 \circ X_2$ which satisfy the following properties:
\begin{itemize}
    \item $X_1$ is independent of $X_2$
    \item $H_{\infty}(X_1) \geq k_1, H_{\infty}(X_2) \geq k_2$
    \item $X_2$ is a space-$s$ source.
\end{itemize}
\EL

Thus we also have the following corollary.

\BCR
For every $s>0$ and every constant $\e>0$ there exists a constant $c>1$ and an explicit extractor $\sext: \zo^n  \to \zo$ for space-$s$ sources with min-entropy $k \geq 2s+c \log n$, and error $\e$.
\ECR

\subsection{Privacy Amplification with an Active Adversary}\label{sec:privacy}
Taking $t=1$ in Theorem~\ref{thm:stnmext}, we get an optimal standard seeded non-malleable extractor.

\BT %\label{thm:stnmext}
 For any constant $\gamma>0$ there is a constant $C>0$ such that for any $0< \e < 1$ with $k \geq C \log(d/\e)$ and $d =C \log(n/\e)$, there is an explicit strong seeded non-malleable extractor for $(n, k)$ sources with seed length $d$, error $\e$ and output length $\frac{(1-\gamma)k}{2}$.  
\ET

Combined with the protocol in \cite{DW09}, we get an optimal two-round privacy amplification protocol with an active adversary.

\BT
There exists a constant $0<\alpha<1$ such that for any $n, k \in \N$, there is an explicit two-round privacy amplification protocol in the presence of an active adversary, that achieves any security parameter $s \leq \alpha k$, entropy loss $O(\log \log n+s)$, and communication complexity $O(\log n+s)$.
\ET

\subsection{Non-Malleable Codes} \label{sec:nmcodes}
Formally, non-malleable codes are defined as follows. 

\BD \cite{ADKO15}
Let $\snm_k$ denote the set of trivial manipulation functions on $k$-bit strings, which consists of the identity function $I(x)=x$ and all constant functions $f_c(x)=c$, where $c \in \bits^k$. Let $E: \bits^k \to \bits^m$ be an efficient randomized \emph{encoding} function, and $D: \bits^m \to \bits^k$ be an efficient deterministic \emph{decoding} function. Let ${\mathcal F}: \bits^m \to \bits^m$ be some class of functions. We say that the pair $(E, D)$ defines an $({\mathcal F}, k, \e)$-\emph{non-malleable code}, if for all $f \in {\mathcal F}$ there exists a probability distribution $G$ over $\snm_k$, such that for all $x \in \bits^k$, we have

\[\left |D(f(E(x)))-G(x) \right | \leq \e.\]  
\ED

\begin{remark}
The above definition is slightly different form the original definition in \cite{DPW10}. However, \cite{ADKO15} shows that the two definitions are equivalent.
\end{remark}

We will mainly be focusing on the following family of tampering functions in this paper.

\BD Let ${\mathcal S}^2_n$ denote the tampering family in the $2$-\emph{split-state-model}, where the adversary applies $2$ arbitrarily correlated functions $h_1,  h_2$ to $2$ separate, $n$-bit parts of string. Each $h_i$ can only be applied to the $i$-th part individually. Let ${\mathcal S}^{affine}$ denote the family of affine tampering functions.
\ED

We remark that in ${\mathcal S}^2_n$, even though the functions $h_1, h_2$ can be correlated, their correlation is independent of the codewords. Thus, they are actually a convex combination of independent functions, applied to each part of the codeword. Therefore, without loss of generality we can assume that each $h_i$ is a deterministic function, which acts on the $i$-th part of the codeword individually.

Cheraghchi and Gursuswami \cite{CG14b} showed that the relaxed two source non-malleable extractor~\ref{def:t2} implies the general definition of non-malleable two-source extractor according to Definition~\ref{def:gnmext} with a small loss in parameters. Specifically, we have

\BL [\cite{CG14b}] \label{lem:nmeq} 
Let $\nmExt$ be a $(k-\log(1/\e), \e)$-non-malleable two-source extractor according to Definition~\ref{def:t2}. Then $\nmExt$ is a $(k, 4\e)$-non-malleable two-source extractor with the general definition.
\EL

Thus, by Theorem~\ref{thm:connection}, Theorem~\ref{thm:nmtext}, and Theorem~\ref{thm:msample2}, we have the following theorem.

\BT
For any $n \in \N$ there exists a non-malleable code with efficient encoding and decoding against $2$-split-state tampering, which has message length $k$, block length $2n$, rate $k/(2n)=\Omega(1)$ and error $2^{-\Omega(k)}$.
\ET

Similarly, Chattopadhyay and Li \cite{ChattopadhyayL17} also showed the relaxed affine non-malleable extractor~\ref{def:anmext} implies the general definition of affine non-malleable extractor according to Definition~\ref{def:gnmext} with a small loss in parameters. Specifically, we have

\begin{lemma}[\cite{ChattopadhyayL17}] \label{lem:affeq}
Let $\nmExt$ be a $(k-\ell, \eps)$-non-malleable extractor for affine sources, according to Definition~\ref{def:anmext}. Then $\nmExt$ is a $(k, \eps+(n+1)2^{-\ell})$-non malleable extractor for affine sources with the general definition.
\end{lemma}

Thus, by Theorem~\ref{thm:connection}, Theorem~\ref{thm:nmaext}, and Theorem~\ref{thm:msample}, we have the following theorem.

\BT
For any $n \in \N$ there exists a non-malleable code with efficient encoding and decoding against affine tampering, which has message length $k$, block length $n$, rate $k/n=\Omega(1)$ and error $2^{-\Omega(k)}$.
\ET

\subsection{Hardness against Read Once Linear Branching Programs}
Chattopadhyay and Liao \cite{ChattopadhyayL22b} showed the following theorem about the hardness against strongly read once linear branching programs.

\BT
Let $\sExt: \zo^n \to \zo$ be a  $(k_1, k_2, \e)$-sumset extractor. Then any strongly read once linear branching program with size at most $2^{n-k_1-k_2-2}$ cannot compute $\sExt$ correctly on more than $\frac{1}{2}+9\e$ fraction of inputs.
\ET

Together with Theorem~\ref{thm:sumsetext}, this gives the following theorem.

\BT
There is an explicit function $\sExt: \zo^n \to \zo$ that requires strongly read once linear branching program of size $2^{n-O(\log n)}$.
\ET
\section{Conclusion and Open Problems} \label{sec:conc}
Our results partially finish several long lines of research projects, which are contributed by numerous researchers and publications. The connections discovered in these projects are amazingly broad. Indeed the techniques that culminated in our main results span areas like pseudorandomness, additive combinatorics, Fourier analysis, cryptography, coding theory and so on. 

There are still interesting and important open problems left. For example, one natural open question is to improve the output length and error of the seedless extractors. Currently for asymptotically optimal entropy, our constructions can only output $1$ bit (or a constant number of bits by the techniques in \cite{Li16}) with constant error, while it is desirable to achieve negligible, or exponentially small error in cryptographic applications. Interestingly, improving the error may also lead to an improvement in output length by the techniques in \cite{Li16}. As observed in previous works, one possible approach is to design $t$-non-malleable extractors with better dependence on $t$, which appears to be a challenging problem. One could also ask if we can construct explicit two-source extractors with entropy $\log n+O(1)$, which would give optimal Ramsey graphs. For non-malleable codes it would be interesting to improve the rates of our codes to optimal. Finally, it is always interesting to find other applications of the pseudorandom objects studied in this paper.

\section{Acknowledgements}
We thank Songtao Mao for pointing out an inaccuracy in an earlier version, and Venkat Guruswami for pointing us to the construction of explicit binary linear codes such that both the code and its dual are asymptotically good in \cite{10.1007/s00037-009-0281-5}.
\bibliographystyle{plain}
\bibliography{refs}
\end{document}